\documentclass[10pt,journal,compsoc]{IEEEtran}
\usepackage{graphicx}
\usepackage{lipsum}
\usepackage{multirow}
\usepackage{dblfloatfix} 
\usepackage[dvipsnames]{xcolor}
\usepackage{soul, url}
\usepackage{color}
\usepackage{boldline}
\usepackage[font=scriptsize]{caption}
\usepackage{adjustbox}
\usepackage{xcolor,colortbl}
\usepackage[noadjust]{cite}

\usepackage[compact]{titlesec}
\usepackage{caption}

\usepackage{tikz}
\def\checkmark{\tikz\fill[scale=0.4](0,.35) -- (.25,0) -- (1,.7) -- (.25,.15) -- cycle;}
\usepackage{pifont}

\newcommand{\xmark}{\text{\ding{55}}}

\definecolor{Gray}{gray}{0.85}
\definecolor{LightCyan}{gray}{0.85}

\newcolumntype{g}{>{\columncolor{Gray}}c}

\usepackage[separate-uncertainty = true,multi-part-units = repeat]{siunitx}
\usepackage{array,amsfonts,amsmath}
\usepackage{subcaption}
\renewcommand{\baselinestretch}{.995}
\usepackage{fancyhdr,geometry}

\begin{document}

\title{Self Supervised Adversarial Domain Adaptation for Cross-Corpus and Cross-Language Speech Emotion Recognition}

\author{Siddique Latif, Rajib Rana, Sara Khalifa, Raja Jurdak~\IEEEmembership{Senior Member,~IEEE}, and Bj\"{o}rn Schuller,~\IEEEmembership{Fellow,~IEEE}
\IEEEcompsocitemizethanks{
\IEEEcompsocthanksitem  S.\ Latif is affiliated with University of Southern Queensland (USQ), Springfield, 4301, Australia and Distributed Sensing Systems Group, Data61, CSIRO, Pullenvale QLD, 4069, Australia. 
\IEEEcompsocthanksitem R. Rana is with University of Southern Queensland (USQ), Springfield, 4301, Australia.
\IEEEcompsocthanksitem  S.\ Khalifa is affiliated with Distributed Sensing Systems Group, Data61, CSIRO, Pullenvale QLD, 4069, Australia.
\IEEEcompsocthanksitem  R.\ Jurdak is affiliated with TruNets -- Trusted Networks Lab,  Queensland University of Technology (QUT), Brisbane, 4000, Australia.
\IEEEcompsocthanksitem  B.\ Schuller is affiliated with GLAM -- the Group on Language, Audio, and Music, Imperial College London, UK, and the ZD.B Chair of Embedded Intelligence for Health Care and Wellbeing, University of Augsburg, Germany.

Corresponding E-mail: siddique.latif@usq.edu.au}}


\IEEEtitleabstractindextext{%
\begin{abstract}

Despite the recent advancement in speech emotion recognition (SER)
within a single corpus setting, the performance of these SER systems degrades significantly for cross-corpus and cross-language scenarios. The key reason is the lack of generalisation in SER systems towards unseen conditions, which causes them to perform poorly in cross-corpus and cross-language settings. Recent studies focus on utilising adversarial methods to learn domain generalised representation for improving cross-corpus and cross-language SER to address this issue. However, many of these methods only focus on cross-corpus SER without addressing the cross-language SER performance degradation due to a larger domain gap between source and target language data. This contribution proposes an adversarial dual discriminator (ADDi) network that uses the three-players adversarial game to learn generalised representations without requiring any target data labels. We also introduce a self-supervised ADDi (sADDi) network that utilises self-supervised pre-training with unlabelled data. We propose synthetic data generation as a pretext task in sADDi, enabling the network to produce emotionally discriminative and domain invariant representations and providing complementary synthetic data to augment the system. The proposed model is rigorously evaluated using five publicly available datasets in three languages and compared with multiple studies on cross-corpus and cross-language SER. Experimental results demonstrate that the proposed model achieves improved performance compared to the state-of-the-art methods.



\end{abstract}

}
\setlength{\headheight}{50pt}

\maketitle \thispagestyle{fancy}

\maketitle

\IEEEdisplaynontitleabstractindextext
\IEEEpeerreviewmaketitle

\IEEEraisesectionheading{\section{Introduction}\label{sec:introduction}}

\IEEEPARstart{S}{peech} Emotion Recognition (SER) is widely explored by researchers to enable effective human-computer interaction. Speech is a major affect display, and it contains information about emotional expressions that can be automatically identified using machine learning (ML) models. SER systems can help businesses by improving their service delivery. Speech emotion identification can be used in call centres to track customer and agent reactions. Speech-based affect recognition can be effectively utilised in healthcare for diagnosis and monitoring of depression, distress, and bipolar disorder in patients \cite{latif2020speech,rana2019automated}. Many other sectors including smart cars \cite{detjen2021increase}, forensic sciences \cite{leshem2020processing}, education \cite{yadegaridehkordi2019affective}, to name a few, are also aiming to utilise SER techniques to improve their performances. 
 
Over the past few years, deep learning (DL) based architectures including deep belief networks (DBN) \cite{hinton2006fast}, convolutional neural networks (CNN) \cite{lecun1989handwritten}, and long short term memory (LSTM) networks \cite{hochreiter1997long} have significantly improved SER performance compared to the classical machine learning (ML) approaches \cite{latif2018transfer,latif2019direct,mao2014learning,latif2020deep}. SER systems based on deep neural networks (DNNs) perform satisfactorily when training and test data belong to the same corpus \cite{latif2020deepdc}. The performance of these systems plummets significantly when the training speech corpus is very different from the testing corpus -- known as cross-corpus SER.

One of the key reasons for poor performance in cross-corpus SER is the difference between training and testing speech data distributions. These differences become more prevalent when training and testing data belong to different languages (cross-language SER). As a solution to the problem, researchers use diverse corpora (including multilingual data) for training to create more generalised and robust SER systems \cite{schuller2010cross}. Studies show that an SER model trained on multiple corpora can achieve improved results \cite{latif2018transfer}. However, acoustic training using multiple labelled data is not feasible for all languages, as we have speech corpora in very few languages compared to the number of languages spoken around the world \cite{latif2021survey,latif2018cross} and getting samples for adaptation in rarely spoken languages is challenging. 


An alternative and more practical approach to address the above challenge is domain adaptation, which generalises SER systems to unseen conditions by minimising domain shift---the gap between source and target data distributions. Domain adaptation approaches maximise the domain confusion to learn a common feature space by minimising some measures of domain shift such as (a) maximum mean discrepancy \cite{long2015learning,yan2017mind} or (b) correlation distances \cite{sun2016return,sun2016deep}. Reconstruction of the target domain using source representation is another way to create a shared representation \cite{ghifary2016deep,ghifary2015domain}. These approaches are effectively used in the computer vision domain. However, achieving domain adaptation in speech emotion is more complex, as it requires keeping the emotional information while reducing domain shift in source and target data.  

Adversarial domain adaptation methods have become a popular manifestation in SER research to minimise an approximate domain discrepancy distance through an adversarial loss. These methods are closely related to the generative adversarial network (GAN) \cite{goodfellow2014generative} training, which pits a generator and a discriminator against each other. The generator is trained to generate fake data in a way that confuses the discriminator. In adversarial domain adaptation, this principle is used among the feature encoder, and domain discriminator \cite{ganin2016domain}. In SER, different studies (e.\,g., \cite{xiao2020learning,yin2020speaker}) use domain discriminator-based adaptation approaches. However, it is difficult to capture all the useful information and complex structures (such as the emotions) in the feature and label spaces using a \textit{single} domain discriminator \cite{jing2021adversarial}.

This paper proposes an Adversarial Dual Discriminator (ADDi) network to learn a domain generalised emotional representation that improves cross-corpus and cross-language SER. Our proposed model is equipped with a dual discriminator, which is not explored in SER. This enables the proposed model to generate the domain invariant representations with a 
three-players adversarial game among generator and dual discriminator. 

To address the challenge of limited labelled data, we further propose self-supervised learning for ADDi -- we call it sADDi. We propose synthetic data generation as our pretext task -- we utilise the unlabelled data to pre-train the encoder component to learn to produce features for emotional synthetic data generation, which can be used to augment the system and help minimise the required labelled training data.

Most of the existing SER studies (e.\,g., \cite{xiao2020learning,yin2020speaker,gideon2019improving}) on adversarial domain adaptation do not consider cross-language, creating a research gap. This is likely due to the complexity of learning a generalised representation for cross-language SER. In this paper, we consider improving cross-language SER.


We summarise the contributions of this paper below. We,
\begin{enumerate}
    \item propose a novel adversarial domain adaptation technique: ADDi for cross-corpus and cross-language SER. ADDi, for the first time, introduces a dual discriminator for SER, enabling the generation of domain invariant representations with a three-players adversarial game among generator and dual discriminator. 
    \item enable self-supervised learning with ADDi (we call it sADDi) by generating synthetic data as a pretext task that effectively utilises unlabelled data to improve the performance and produce synthetic data to augment the SER system. 
    \item use five widely applied publicly available datasets to comprehensively evaluate the performance of ADDi and sADDi for cross-corpus and cross-language SER performance. Results show that ADDi outperforms the state-of-the-art methods and when including self-supervised learning in ADDi, sADDi offers even higher performance improvement than the state-of-the-art methods. Besides improving the performance, sADDi reduces the required amount of source labelled data by 15-20\,\% compared to the current and most relevant study (ADDoG) \cite{gideon2019improving} while achieving comparable classification accuracy. 
\end{enumerate}

\section{Related Work}

\begin{table*}[]
\caption{Summary of a comparative analysis of our paper with that of the existing literature. Very few studies have used Adversarial Learning for cross-language SER. (checked \color{green}{green})}
\label{table:comparison}
\centering
\begin{tabular}{|l|l|l|l|l|l|}
\hline
  &         & \multicolumn{2}{c|}{Evaluations} &
  & \cellcolor[HTML]{C0C0C0}  \\ \cline{3-4}
  \multirow{-2}{*}{Author (Year)} 
  & \multirow{-2}{*}{\begin{tabular}[c]{@{}l@{}}Technique\end{tabular}} 
  & Cross-Corpus   & Cross-Language  
  & \multirow{-2}{*}{\begin{tabular}[c]{@{}l@{}}Adversarial \\ Learning\end{tabular}} 
  & \multirow{-2}{*}{\cellcolor[HTML]{C0C0C0}\begin{tabular}[c]{@{}l@{}}Self-Supervised\\ Learning\end{tabular}} \\ \hline

 Schuller et al. (2010) \cite{schuller2010cross}&\begin{tabular}[c]{@{}l@{}}Feature\\normalisation\end{tabular}       &    \checkmark             &   \checkmark             &  \xmark                                              & \xmark \cellcolor[HTML]{C0C0C0}\\\hline 
 Zhang et al. (2011) \cite{zhang2011unsupervised}&\begin{tabular}[c]{@{}l@{}}Feature\\normalisation\end{tabular}       &    \checkmark             &  \checkmark            &  \xmark                                              & \xmark \cellcolor[HTML]{C0C0C0}\\\hline 
  
Kim et al. (2017) \cite{kim2017towards}&\begin{tabular}[c]{@{}l@{}}Aggregated corpora\\training\end{tabular}       &    \checkmark             &  \checkmark            &  \xmark                                              & \xmark \cellcolor[HTML]{C0C0C0}\\\hline

Latif et al. (2018) \cite{latif2018transfer}&\begin{tabular}[c]{@{}l@{}}Transfer learning\end{tabular}       &    \checkmark             &  \checkmark            &  \xmark                                              & \xmark \cellcolor[HTML]{C0C0C0}\\\hline

Neumann et al. (2018) \cite{neumann2018cross}&\begin{tabular}[c]{@{}l@{}}Aggregated corpora\\training\end{tabular}       &    \checkmark             &  \checkmark            &  \xmark                                              & \xmark \cellcolor[HTML]{C0C0C0}\\\hline

Abdelwahab et al. (2018) \cite{abdelwahab2018domain}&Domain adaptation         &    \checkmark             &   \xmark              &  \checkmark                                             & \xmark \cellcolor[HTML]{C0C0C0}\\\hline


Latif et.al \cite{latif2019unsupervised} (2019) &    Domain adaptation                                                                            &  \xmark              & \color{green} {\checkmark}               &  \color{green} {\checkmark}      & \xmark \cellcolor[HTML]{C0C0C0}  \\ \hline

Song et.al \cite{song2019transfer} (2019) & \begin{tabular}[c]{@{}l@{}} Feature subspace\\learning  \end{tabular}                                                                         &  \checkmark             &  \checkmark               & \xmark        & \xmark \cellcolor[HTML]{C0C0C0}  \\ \hline

Gideon et al. (2019) \cite{gideon2019improving}& Domain adaptation                    &    \checkmark   &    \xmark & \checkmark   & \xmark \cellcolor[HTML]{C0C0C0}  \\ \hline

Xiao et al. (2020) \cite{xiao2020learning}& Domain adaptation                    &    \checkmark   &    \xmark & \checkmark   & \xmark \cellcolor[HTML]{C0C0C0}  \\ \hline


Luo et.al \cite{luo2020nonnegative} (2019) & \begin{tabular}[c]{@{}l@{}} Feature subspace\\learning  \end{tabular}                                                                         &  \checkmark             &  \checkmark               & \xmark        & \xmark \cellcolor[HTML]{C0C0C0}  \\ \hline

Ahn et al. (2021) \cite{ahn2021cross}&\begin{tabular}[c]{@{}l@{}} Domain adaptation \& \\aggregated corpora\\training  \end{tabular}                   &    \checkmark   &    \color{green} {\checkmark} & \color{green} {\checkmark}   & \xmark \cellcolor[HTML]{C0C0C0}  \\ \hline


\rowcolor[HTML]{C0C0C0} Ours (2022) &Domain adaptation &    \checkmark            &       \checkmark          &  \checkmark                                                                                 &   \checkmark

                                \\ \hline
\end{tabular}
\label{litrature}
\end{table*}

\subsection{Cross-Corpus and Cross-language SER}
Cross-corpus speech emotion recognition is an important task to enable real-life SER applications. It aims to build systems with improved generalisation to perform SER not only in variations in speaker and languages but also in unknown target conditions, including changes in recording environments, noise levels, and elicitation strategy. State-of-the-art SER systems trained on a single corpus fail to perform well in cross-corpus settings. Previous studies explore various techniques to achieve better performance in cross-corpus SER. Schuller et al.\  \cite{schuller2010cross} find that the SER performance degrades due to the acoustic and annotation differences. They perform experiments using six corpora to gain generalisation. They also evaluate multiple normalisation techniques and z-normalisation to achieve the best results. Eyben et al.\  \cite{eyben2010cross} perform cross-corpus SER evaluations using speech databases with realistic and non-prompted emotions. They use a uni-variate ranking of the low-level descriptors (LLDs) to find the most important features and achieve improvement in some settings. They highlight that future efforts are required to address the inconsistencies among multiple corpora by carefully selecting annotations. Zhang et al.\  \cite{zhang2011unsupervised} evaluate unsupervised learning and feature normalisation for cross-corpus SER. They show that 
adding 
unlabelled data to agglomerate multi-corpus training sets and utterance level feature normalisation can improve performance. In \cite{schuller2011using}, the authors show the effect of data agglomeration and decision-level fusion for cross-corpus SER. They use six datasets and demonstrate that joint training with multiple corpora and late fusion could help improve performance. These studies show the preliminary feasibility of cross-corpus learning and motivate further in-depth research. 

Researchers also explore different techniques to perform emotion identification in cross-language settings. Albornoz et al.\   \cite{albornoz2015emotion} consider emotion profile-based ensemble support vector machines (SVM) for emotion classification in multiple languages. They model each language independently to preserve the cultural properties and apply the universality of emotions to map and predict emotions in different languages. They use the RML corpus \cite{wang2008recognizing} and achieve improved results using their model in a language-independent SER. Li et al.\ \cite{li2019improving} develop a three-layered model of acoustic features, semantic primitives, and emotion dimensions to perform cross-language emotion classification. They apply feature selection and speaker normalisation and evaluate the proposed framework on Japanese, German, Chinese, and English emotional speech corpora. They achieve multilingual recognition performance comparable with a monolingual emotion recogniser. In \cite{latif2018cross}, the authors evaluate cross-lingual SER and highlight the ways of designing an adaptive emotion recognition system for languages with a small available dataset. They show that training the model with multiple languages data can deliver comparable results with a model trained with monolingual data and that augmentation of the training set with a fraction of target language labelled data can help improve the performance. Various other studies (e.\,g., \cite{schuller2011using,feraru2015cross,eyben2010cross}) explore cross-lingual SER, however, these studies evaluate classical ML models on relatively smaller datasets. 

Most recent studies on SER utilise deep representation learning techniques over low-level features. Particularly, studies use deep networks to learn generalised representations to improve performance. For instance, the authors in \cite{latif2018transfer} use DBNs for learning generalised features across multiple datasets. They evaluate the proposed model using six emotional corpora and showed that DBN can provide better performance in cross-corpus SER. They also observe that a DBN can learn a robust representation from many language datasets that helps improve SER performance. In \cite{neumann2018cross}, the authors train an attentive convolutional neural network (ACNN) for binary classification of arousal and valence in cross-language and multi-language training settings using French and English language datasets. They show that multilingual training can enhance the performance of the system. Also, they find that the ACNN can be fine-tuned using a fraction of target language data to produce sound results for cross-language SER. Ning et al.\  \cite{ning2017learning} employ multilingual Bidirectional Long Short-Term Memory (BLSTM) with the shared hidden layers across different languages for universal feature representation learning. They evaluate the proposed model for English and Mandarin corpora and found that cross-lingual knowledge learning using shared hidden layers helps improve SER performance compared to BLSTM variants without shared hidden layers. Some other studies (e.\,g., \cite{zhang2017multi,parry2019analysis,kim17d_interspeech}) also exploit deep networks to improve cross-corpus and cross-language emotion detection. In general, the methods proposed in these studies require large aggregated speech labelled corpora to achieve generalisation for improved cross-corpus performance. Models training using aggregated corpora is not feasible in real life, as it requires multiple labelled datasets. In contrast, domain adaptation is a more practical approach that improves the system's generalisation without the need for multiple labelled corpora. We review the studies on domain adaptation in the next subsection.

\subsection{Adversarial Domain Adaptation}
Deep domain adaptation aims to improve the generalisation of SER systems by addressing the problem of domain shift among source and target datasets. Researchers explore different domain adaptation models (e.\,g., \cite{deng2014autoencoder,deng2017universum,abdelwahab2015supervised,ocquaye2019dual}) to improve cross-corpus and cross-lingual SER.  To this end, adversarial domain adaptation techniques are becoming very popular in SER. For instance, the authors in \cite{abdelwahab2018domain} use domain adversarial neural networks (DANN) \cite{ganin2016domain} for cross-corpus emotional attributes' prediction.
They learn generalised representations between the source and target data by using a gradient reversal layer (GRL) which propagates back the negative of the gradient produced by the domain classifier to the shared network. They find that the DANN can learn domain invariant
representations to cross-corpus SER. Xiao et al.\ \cite{xiao2020learning} propose an adversarial network for class-aligned
and generalised domain adaptation. They also exploit GRL to generalise representations among source and target data. They evaluate the proposed model against cross-corpus settings using IEMOCAP and MSP-IMPROV corpora and achieved improved results compared to DANN and AE-based deep architectures. Zhou et al.\ \cite{zhou2019transferable} present a class-wise domain adversarial adaptation method to learn common representation to address cross-corpus mismatch issues. They evaluated the proposed model on two datasets including AIBO and EMO-DB for the French language and show that the proposed model achieves better results when training is performed on target data with minimal labels for positive and negative emotion classes recognition. Gideon et al.\  \cite{gideon2019improving} introduce an adversarial discriminative domain generalisation model that follows a ``meet in the middle'' approach for cross-corpus emotion recognition. The proposed approach utilises the critic network that enables the model to improve the cross-corpus generalisation by iteratively moving representations closer to source and target data. They perform evaluations using English datasets including IEMOCAP, MSP-IMPROV, and PRIORI emotion datasets \cite{gideon2019improving} and show that the proposed framework generates generalised representations for improved cross-corpus SER.

Most of the studies above evaluate adversarial domain adaptation methods for cross-corpus SER using similar language corpora. However, few studies show the effectiveness of their methods for different languages in cross-corpus SER. Ahn et al.\ \cite{ahn2021cross} propose a few shots learning-based unsupervised domain adaptation techniques to learn emotional similarity among source and target domains. They evaluate the proposed model in three different languages and achieve improved results. However, their proposed method requires additional labelled training data to improve the generalisation. Latif et al.\ \cite{latif2019unsupervised}  present a GAN-based adversarial method to learn language invariant representations and evaluate the model for different language datasets. They train support vector machines (SVM) on language invariant representations to improve the performance of cross-language SER. In contrast to these studies, we propose an Adversarial Dual Discriminator (ADDi) network that utilises a dual discriminator to learn generalised representations to improve cross-corpus SER. One of the novel features of our model is the utilisation of self-supervised learning (SSL) for domain adaptation, which has not been explored for SER domain adaptation. Few studies exploited SSL for improving SER performance within-corpus settings. We discuss these studies in the following subsection.

\subsection{Self-Supervised SER}

Self-supervised learning (SSL) \cite{raina2007self} is a new paradigm in ML, which uses data for supervision. The self-supervised task, also known as the pretext task, uses the unlabelled data to guide downstream tasks. SSL-based models are getting tremendous interest in computer vision \cite{jing2020self}, natural language processing (NLP) \cite{lan2019albert}, and automatic speech recognition (SER) \cite{baevski2019vq}; however,  few studies utilise SSL in SER.  In \cite{pascual2019learning}, the authors propose a multitask SSL technique to learn a shared speech representation, where a single encoder network is followed by multiple workers that jointly solve different self-supervised tasks. They perform evaluations on speaker, phoneme, and emotional cue recognition, and achieve improved results. Self-supervised multi-modal representation learning though transformers \cite{vaswani2017attention} is increasingly gaining momentum to improve SER \cite{macaryuse}. Khare et al.\ \cite{khare2020self} use transformer-based SSL to improve the performance of multimodal emotion recognition. They fine-tune a transformer trained on a masked language modelling task and can improve emotion recognition performance by 3\,\% on the CMU-MOSE dataset \cite{zadeh2018multimodal}. A recent study \cite{shukla2020visually} presents a visually-guided SSL framework for improving the SER performance. The authors generate video frames using still images by conditioning the network on corresponding audio. In this way, the pre-trained encoder part of their network learns important features to generate realistic facial and lip movements. They hypothesise that the features learnt by the encoder are highly correlated with the presence of emotion and particular phonemes. They utilise these representations for ASR and SER to achieve state-of-the-art results. In contrast, we propose to generate synthetic emotional data as a pretext task, which adversarially enables the encoder to encode discriminative features for emotional data generations. We use this encoder for our downstream domain adaptation task, which helps produce emotionally discriminative features while minimising the gap between source and target domains. In addition, the synthetic emotional data generated in our downstream task acts as a by-product that can be utilised to augment the system. 

\subsection{The Research Gap (Summary)}
The related work can be summarised as follows.
\begin{itemize}
    \item Several studies show that DL models trained using multiple sources corpora can improve cross-corpus SER performance; however, acoustic training from multiple language data in real-life is not a feasible approach due to the unavailability of sufficient labelled data for multiple languages. Therefore, there is a need for new methods to overcome this limitation. 
    \item Adversarial neural networks based domain adaptation approaches are widely used for cross-corpus SER; however, there is still room for performance improvement, particularly for cross-language SER. 
     \item Self-supervised learning can be used as an effective tool to address the limited label issue but has not been fully explored and used for cross-corpus SER. 
\end{itemize}
In Table~\ref{table:comparison}, we contrast our work with the literature, briefly showcasing how we aim to address the research gaps.

\begin{figure*}[!t]%
\centering
\includegraphics[trim=0cm 0cm 0cm 0cm,clip=true,width=0.7\linewidth]{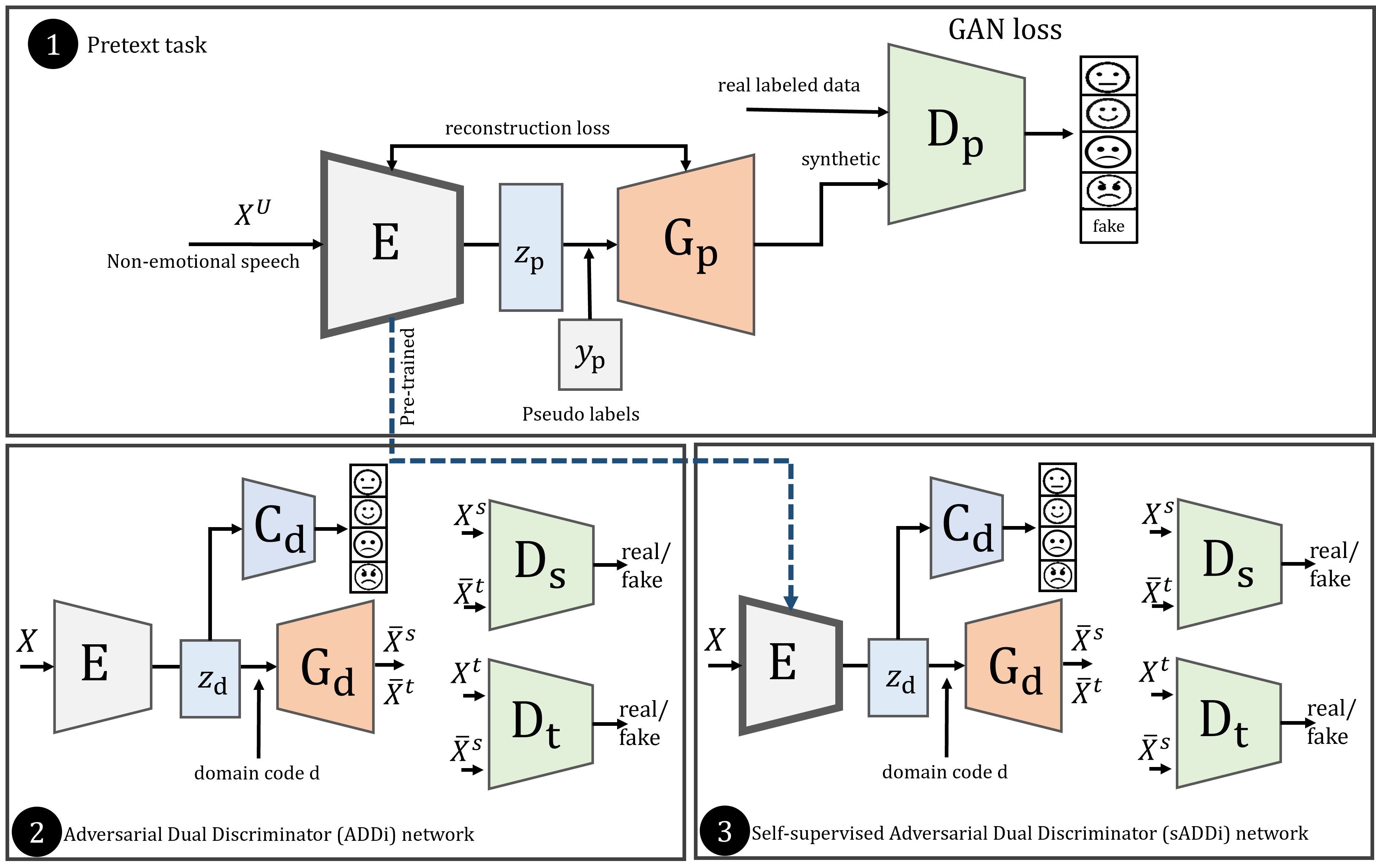}%
\captionsetup{justification=centering}
\caption{Overall structure of the proposed framework. We use numbers for different components, where (1) shows the pretext task that is trained on synthetic data generation; (2) represents the Adversarial Dual Discriminator (ADDi) for domain adaptation; and (3) shows the self-supervised Adversarial Dual Discriminator (sADDi), where we use pre-trained encoder from the pretext task as highlighted with a dashed line.}%
\label{model}%
\end{figure*}

\section{Proposed Model}
\label{Adver}

The core of the proposed model is the Adversarial Dual Discriminator (ADDi) network and the module for the pretext task enabling self-supervision. We propose the generation of synthetic data as a pretext task, wherein we essentially pre-train an encoder that we later use to realise the proposed Self-supervised Adversarial Dual Discriminator (sADDi) network.

\subsection{Adversarial Dual Discriminator (ADDi) network}
 Our proposed Adversarial Dual Discriminator (ADDi) network is equipped with an encoder ($E$), a generator ($G_p$), and dual discriminators: ($D^s$) and ($D^t$). An overview of the proposed framework is shown in Figure \ref{model}, where subfigure with label 2 shows ADDi network. It performs adversarial domain adaptation by learning domain invariant features.  We represent the source domain data and target domain data as $X^s=\{x_{i}^{s},y_{i}^s\}_{i=1}^n$ and $X^t=\{x_{j}^t\}_{j=1}^{m}$, respectively. The encoder ($E$) attempts to map the input data either from source or target to a domain invariant latent representation ($z_d=E(X), X \in X^s \cup X^t$). The generator ($G_d$) conditioned on domain code $d$ uses this domain invariant latent representation ($z_d$) to generate source $\bar{X}^s$  or target $\bar{X}^t$ domain samples. The generator $G_d$ is adversarially connected with two discriminators.  
The objective function of the generator is as follows:
\begin{equation}
  \mathcal{L_\text{G}} = \mathcal{L}_{\text{AE}} + \lambda\mathcal{L}_{G}^{\text{adv}},
\end{equation}
\begin{equation}
\begin{aligned}
\mathcal{L}_{G}^{\text{adv}} =&\mathbb {E}[\log(1 - D_{s}(G_d(E(X),d)))] \quad\\+&\mathbb {E}[\log(1 - D_{t}(G_d(E(X),d)))],
\end{aligned}
\end{equation}

\begin{equation}
\label{AE}
    \mathcal{L_{\text{AE}}}(X,G_{d}(E(X)))=\lVert{X-\bar{X}}\rVert_{2}^{2}, 
\end{equation}
where $\lambda$ is a balancing parameter. The generator is connected to the dual discriminators $D_s$ and $D_t$, which play the three-players minimax adversarial game to produce $z_d$ to be domain invariant. The generator also acts as the decoder and it reconstructs back the input samples $X$ with the latent representation $z_d$ using the reconstruction loss in Equation \eqref{AE}. 

The dual discriminators are tasked to distinguish the real data from the fake data. Particularly, for domain code $d=0$, the discriminator $D^s$ differentiates between $ \tilde{X}^t=G_{p}(E(X),d)$ (fake) and source data $X^s$ (real), whereas the discriminator $D^t$ discriminates between $ \tilde{X}^s=G_{p}(E(X),d)$ (fake) and source data $X^s$ (real), for domain code $d=1$. The adversarial process of the generator $G_{p}$ and two discriminators minimises the divergence between the source and target data distributions and forces the encoder $E$ to generate a generalised latent representation $z_d$ across the source and target domains. The objective function for the dual discriminators can be given as follows:

\begin{align*} \mathcal {L}_{ D_{s}}=&\mathbb {E}\left [{-{\mathrm{ log}} D_{s}\left ({X^s}\right)}\right] \!+\!\mathbb {E}\left [{-{\mathrm{ log}}\left ({1- D_{s}\left ({G_d\left ({X,d}\right)}\right)}\right)}\right]_{\{d=0\}}, \qquad ~~\tag{11}\\
\mathcal {L}_{ D_{t}}=&\mathbb {E}\left [{-{\mathrm{ log}} D_{t}\left ({X^t}\right)}\right] +\mathbb {E}\!\left [{-{\mathrm{ log}}\left ({1\!-\! D_{t}\left ({ G_d \left ({X,d}\right)}\right)}\right)\!}\right]_{\{d=1\}}. \tag{12}\end{align*}

The classifier $C_d$ is connected with the latent representation $z_d$ and minimises the cross entropy loss for emotion classification during training using only the source data labels and the error is back-propagated through the network to update $E$. In this way, the encoder $E$ gets influenced by the classifier and enforces $z_d$ to be emotionally discriminating as representation. This helps produce emotionally discriminative and domain invariant representations to perform cross-corpus and cross-language robust SER. When the pre-trained encoder $E$ is fine-tuned in the domain adaptation task, it promotes the discriminative power of a domain invariant representation and further boosts the performance of the system. We discuses the encoder pre-training in the next section. 

During training, first, the autoencoder is updated using the equation \eqref{AE}. Afterwards, $G_d$ is updated to generate the fake samples using $z_d$ and the domain code $d$. We concatenate the one-hot domain code with the encoded representation $z_d$ and feed to $G_d$. We further update the discriminators based on the domain codes. For the samples with $d=0$, $D^s$ is updated, whereas $D^t$ is updated for the samples with $d=1$. Finally, we update the $C_d$ for the source data samples.

 \subsection{Pretext: Synthetic Emotional Data Generation and Self Supervised Adversarial Dual Discriminator (sADDi)}
A pretext task is used in self-supervised learning (SSL) to generate useful representations that can provide a supervisory signal to the down stream task. It is a predefined task for the network to solve learning the objective function \cite{jing2020self}.  Most of the SSL pretext tasks are designed based on intuition or heuristics \cite{xiao2021self}. There is no guarantee on the compatibility between the pretext task and the down stream task \cite{zhang2017split}. For SER, solving multiple audio based self-supervised tasks can offer improvements \cite{pascual2019learning}. However, these tasks have been evaluated for within corpus SER settings. The design of an SSL pretext task for domain adaption is challenging, as emotionally discriminative generalised representations are required to effectively perform cross-corpus SER. 

We use synthetic emotional data generation as pretext task for cross-corpus domain adaptation. The intuition here is that the encoder network pre-trained to encode discriminative features for emotional synthetic data generation when utilised in domain adaptation should help produce an emotionally discriminative generalised representation. 

The architecture for our pretext task is shown in Figure \ref{model} as a subfigure with label 1. It follows the GAN architecture consisting of a generator and a discriminator. Both these networks play an adversarial game defined by the following optimisation program in Equation \ref{eq:gan}. 
 \begin{equation}
 \label{eq:gan}
 \underset {G}{\min }\,\underset {D}{\max }\, \mathbb {E}_{x\sim P_{\text {data}}}\left [{{\mathrm{ log}}D\left ({x}\right)}\right]+\mathbb {E}_{x\sim P_{G}}\left [{{\mathrm{ log}}\left ({1-D (G(z))}\right)}\right].   
 \end{equation}
 The generator network captures the data distribution and generates new samples by incorporating feedback from the discriminator network. The discriminator network in a GAN is simply a classifier. It tries to classify the real and fake data, generated by the generator network. While there are many variants of GAN architectures (e.\,g., \cite{choi2018stargan,arjovsky2017wasserstein,brock2018large}), we use the balancing GAN \cite{mariani2018bagan} like architecture due to its effectiveness in SER \cite{eskimez2020gan}. It consists of an encoder ($E$), generator/decoder ($G_p$), and discriminator ($D_p$). The encoder network ($E$) takes the non-emotional speech data ($X^U$) and generates latent code $z_p$. We concatenate the $z_p$ with the pseudo labels ($y_p$) and feed to the generator ($G_p$) to generate the synthetic data. Since the unlabelled samples do not belong to any emotional class, the pseudo labels in four classes are randomly generated and uniformly distributed to the unlabelled non-emotional speech. In this way,  the $G_p$ network conditioned on the ($y_p$) has explicit emotion class label information during generation like the conditional GAN \cite{mirza2014conditional}. During the adversarial training, $G_p$ is tasked to generate samples in different classes based on $y_p$; and $D_p$ is trained to differentiate the generated samples (by generator ($G_p$)) as fake and real samples to their class labels. The generator tries to avoid the fake label and matches the desired emotional class labels. The discriminator is optimised to output $Nc+1$ neurons, where $N_c$ represents the emotional classes (happy, sad, neutral, or angry) and the last neuron represents the fake class as used in \cite{mariani2018bagan,chatziagapi2019data}. Since the encoder ($E$) is coupled with the GAN, it learns to encode features for different emotional classes in the latent space of the generator $G_p$. After pre-training, we fine-tune the encoder network in our ADDi network which we name self-supervised Adversarial Adversarial Dual Discriminator (sADDi). It is highlighted in the Figure \ref{model} with blue dashed line.

\section{Experimental Setup}

\subsection{Datasets}
\label{sec:data}
To evaluate the performance of our proposed model, we use five different emotional datasets, including IEMOCAP, MSP-IMPROV, RECOLA, EMODB, and FAU-AIBO, which are commonly used for cross-corpus and cross-language emotion classification research \cite{latif2021survey}. In order to use additional unlabelled data for self-supervised learning (SSL), we use a subset of Librispeech~\cite{panayotov2015librispeech}, which is a corpus of read English speech, suitable for training and evaluating models on automatic speech and speaker recognition systems. Below, we briefly describe these datasets.

\subsubsection{IEMOCAP}
This database contains 12 hours of audiovisual data, including audio, video, textual transcriptions, and facial motion information \cite{busso2008iemocap}. The recordings are collected from $10$ professional actors (five males and five females) during dyadic interactions. In contrast to reading text with prototypical emotions, dyadic interactions allowed the actors to perform more spontaneous emotion \cite{lotfian2017building}. For categorical labels, each sentence is annotated by three annotators and the participant. Finally, an utterance is assigned a label if at least three annotators are assigned the same label. Overall, IEMOCAP contains nine emotions: excited, happy, sad, neutral, angry, disgust, frustrated, fearful, and surprised. Similarly to previous studies \cite{latif2018variational}, we only use utterances of four categorical emotions, including happy, neutral, sad, and angry in this study by merging ``happy'' and ``excited'' as one emotion class ``happy''. The final dataset includes $5\,531$ utterances ($1\,636$ happy, $1\,708$ neutral, $1\,084$ sad, and $1\,103$ angry instances). For continuous labels, IEMOCAP is also annotated for arousal and valence on a scale of 1 to 5. We map continuous labels to binary labels as presented in Table \ref{table:mapping}.    
\subsubsection{MSP-IMPROV}
The MSP-IMPROV dataset is an acted audiovisual emotional database recorded from $12$ speakers performing dyadic interactions \cite{busso2017msp}. Overall, the recordings are grouped into six sessions and each session contains the recordings of one male, and one female actor similar to IEMOCAP \cite{busso2008iemocap}. The scenarios were carefully designed to control emotion and lexical content while maintaining naturalness in the recordings. The MSP-IMPROV is annotated through perceptual evaluations using crowdsourcing \cite{burmania2016increasing}. This corpus contains utterances in four categorical emotions: angry, happy, neutral, and sad. To be consistent with previous studies \cite{latif2019direct,gideon2017progressive},  we use all utterances with four emotions: anger (792), happy (2\,644), sad (885), and neutral (3\,477).
\subsubsection{RECOLA}
RECOLA \cite{ringeval2013introducing} is a French multimodal corpus of spontaneous collaborative and affective interactions. While solving a collaborative task, speakers recorded the dyadic conversations during a video conference. 46 participants (27 females, and 19 males) were recruited to record this corpus. We use the publicly available portion of RECOLA, which contains $1,308$ utterances of $23$ speakers. An open-source web-based tool ANNEMO\footnote{https://diuf.unifr.ch/main/diva/recola/annemo} was developed for its affective annotation. RECOLA is annotated with continuous labels, including arousal and valence in the range $[-1, 1]$. We use RECOLA for cross-corpus language SER and perform binary classification of arousal
(low/high) and valence (negative/positive) as considered in \cite{neumann2018cross}. Table \ref{table:mapping} shows the mapping of original annotations to a binary scheme for IEMOCAP and RECOLA. 

\begin{table}[!ht]
\centering
\caption{Mapping Rules for IEMOCAP and RECOLA. Here bracket includes the elements listed and the parenthesis does not contain the listed elements. }
\begin{tabular}{lcc}
\hline
\textbf{Corpus}              & \multicolumn{1}{l}{\textbf{Low/Negative}} & \multicolumn{1}{l}{\textbf{High/Positive}} \\ \hline
\multicolumn{1}{l|}{IEMOCAP} & \multicolumn{1}{c|}{{[}1, 2.5{]}}         & (2.5, 5{]}                                 \\ \hline
\multicolumn{1}{l|}{RECOLA}  & \multicolumn{1}{c|}{{[}-1, 0{]}}          & (0, 1{]}                                   \\ \hline
\end{tabular}
\label{table:mapping}
\end{table}

\subsubsection{EMODB}

EMODB \cite{burkhardt2005database} is the most popular and widely used publicly available emotional dataset in German Language, recorded by the Institute of Communication Science, Technical University Berlin. It contains audio recordings of seven emotions recorded by ten professional speakers in 10 German sentences. This study selects four basic emotions: happy, sad, neutral, and angry to perform categorical cross-language emotion recognition.

\subsubsection{FAU-AIBO}
FAU-AIBO \cite{steidl2009automatic} corpus is a spontaneous emotional corpus in the German language. It contains 9.2 hours of speech from 51 children from different schools while interacting with Sony's pet robot AIBO. In this study, we select FAU-AIBO to evaluate the proposed framework against completely naturalist emotional speech. We map this corpus to binary valence for evaluations.

\subsubsection{LibriSpeech}
The LibriSpeech dataset \cite{panayotov2015librispeech} contains $1\,000$ hours of English read speech from $2\,484$ speakers. This corpus is derived from audiobooks and is commonly used for automatic speech and speaker recognition problems \cite{berard2018end,dubey2019transfer}. The training portion of LibriSpeech is divided into three subsets, with an approximate recording time of $100$, $360$ and $500$ hours. This paper uses the subset that contains $100$ hours of recordings. These recordings span over $251$ speakers. 

\subsection{Speech Features Extraction}
We represent the speech sample in Mel Filter Banks (MFBs), a widely used speech representation in speech research \cite{xu2017convolutional,gideon2019improving}. We use the Kaldi speech recognition toolkit \cite{povey2011kaldi} to extract $40$-dimensional
MFBs from each utterance. To extract MFBs, we use default options, including a Povey window with a frame length of $25$\,ms and a frameshift of $10$\,ms, a preemphasis coefficient of $0.97$, and a low cutoff of $20$\,Hz. These configurations are selected based on \cite{gideon2019improving} to make a fair comparison. Due to the varying lengths of the audio samples, we pad the MFBs with zeros to the length of the longest emotional utterance.

\subsection{Model Configuration}
 This subsection presents the configuration of three models, including a baseline Convolutional Neural Network (CNN), the proposed Adversarial Dual Discriminator (ADDi) network, and the pretext task GAN. Each of these models takes MFBs as the input feature set. Each experiment considers labelled source data for training, and target data is used for testing. We train all these models using Adam as the optimiser with default parameters and a starting learning rate of $0.0001$. We compute the validation accuracy at the end of each epoch during training. If the validation accuracy did not improve after 5 epochs, we restore the model to the best epoch and halves the learning rate. This process continues until the learning rate reaches below $0.00001$. 
 
 The CNN baseline network comprises a feature encoder and emotion classifier. The feature encoder consists of convolutional and max-pooling layers, whereas the classifier part utilises the fully connected layers for classification. Due to the unavailability of target data in the experiments, it is difficult to validate all the hyperparameters of the network for cross-dataset SER. Therefore, we select the parameters commonly used in prior studies \cite{latif2020augmenting,latif2020multi,gideon2019improving}. The feature encoder has three convolutional layers, each followed by the pooling layers. We start with a large filter size of $16$ in the first convolutional layer as suggested by prior work \cite{latif2020multi}. The convolutional layers capture the salient regions within the MFBs and create the feature maps. The pooling layers reduce the dimension of these feature maps by identifying the most relevant features. We use the max-pooling layer to give better performance than average pooling during experiments. The feature encoder encodes the entire utterances into the $256$ features. The classifier uses these features for emotion classification. We have two dense layers with hidden units of $256$ and $128$. We employ a dropout layer between two dense layers with a dropout rate of $0.3$ to avoid overfitting. 
 
The ADDi network also has an encoder component to encode input MFBs to the domain invariant representation that is used by the generator. We apply a similar encoder architecture to the baseline CNN. The decoder/generator has three transposed convolutional layers to generate samples using the encoded latent representation. Two discriminators and the classifier of ADDi have two hidden layers containing hidden units of $256$ and $128$ in number. Like the baseline CNN and ADDi, our pre-training GAN also has an encoder network that follows a similar architecture. We employ the same architecture for the discriminator as for the ADDi network. We select the Rectified Linear Unit (ReLU) as a non-linear activation function for all the models due to its better performance than hyperbolic tangent and leaky ReLU during validation.

\section{Experiments and Evaluations}

We apply a two-tiered evaluation approach: We evaluate ADDi to understand the significance of the proposed dual-discriminator based framework. We then evaluate sADDi to understand the relative significance of self-supervised learning for ADDi. We evaluate the performance of the proposed ADDi and sADDi networks in cross-corpus and cross-language settings by comparing them with related studies that report similar results. To further extend the extent of our comparison, we implement related models including a CNN (baseline), a GAN \cite{latif2019unsupervised}, a DANN \cite{abdelwahab2018domain}, a DBN \cite{latif2018transfer}, a CNN-LSTM \cite{parry2019analysis}, and an autoencoder-based model as used in \cite{deng2013sparse} and compare our results with these models. We repeat each experiment ten time and calculated mean and standard deviation. Results are presented using the unweighted average recall rate (UAR), a widely accepted metric in the field.

\subsection{Cross-Corpus Results}
We evaluate the proposed ADDi and sADDi networks for cross-corpus SER using the IEMOCAP and MSP-IMPROV datasets. Both of these datasets are recorded in similar laboratory conditions in English. In this experiment, we consider no labels for the target dataset. We use a random 80:20 (train:test) split of the source data and train the model as used in \cite{gideon2019improving}. 
We compare the performance of the ADDi network with a baseline CNN, ADDOG \cite{gideon2019improving}, a DANN and a GAN~\cite{latif2019unsupervised}. The results are presented in Table \ref{crosscorpus}.

\begin{table}[!ht]
\caption{Cross-corpus SER results in UAR ($\%$) using IEMOCAP (English) and MSP-IMPROV (English).}
\centering
\scriptsize
\begin{tabular}{|l|l|l|}
\hline
Model    & \multicolumn{1}{c|}{\begin{tabular}[c]{@{}c@{}}IEMOCAP to \\ MSP-IMPROV\end{tabular}} & \multicolumn{1}{c|}{\begin{tabular}[c]{@{}c@{}}MSP-IMPROV to\\ IEMOCAP\end{tabular}} \\ \hline
CNN \scriptsize{(baseline)}      &42.5$\pm$1.6     & 44.3$\pm$1.5\\ \hline
DANN \cite{abdelwahab2018domain}&42.8$\pm$1.4&44.9$\pm$1.7\\\hline
GAN \cite{latif2019unsupervised}& 43.6$\pm$1.3   &45.8$\pm$1.5\\\hline
ADDOG \cite{gideon2019improving} & 44.4$\pm$0.9   &47.4$\pm$0.7\\\hline
\rowcolor[HTML]{EFEFEF}
ADDi \scriptsize{(proposed)} & \textbf{45.1$\pm$0.8}       & \textbf{48.2$\pm$0.6}         \\ \hlineB{3}\hlineB{1}
CNN$_\text{SSL}$ \scriptsize{(baseline)}      & 43.8$\pm$1.2                & 45.3 $\pm$ 1.1                                                                             \\ \hline
\rowcolor[HTML]{EFEFEF}
sADDi \scriptsize{(proposed)} & \textbf{47.1$\pm$0.5}                      & \textbf{49.8$\pm$0.6}                                                                              \\ \hlineB{3}\hlineB{1}
\end{tabular}
\label{crosscorpus}
\end{table}

Compared to these existing methods and baseline, ADDi achieves better results. ADDi achieves $2.6 \%$ and $3.9\%$ relative improvements compared to the baseline CNN for IEMOCAP to MSP-IMPROV and MSP-IMPROV to IEMOCAP experiments, respectively.  Amongst the previous studies, ADDOG utilises the critic component similar to a Wasserstein GAN \cite{arjovsky2017wasserstein} to learn generalised representations for cross-corpus SER, while another study \cite{latif2019unsupervised} applies a single discriminator based adversarial method to minimise the domain gap, and whereas in \cite{abdelwahab2018domain}, a gradient reversal layer (GRL) \cite{ganin2016domain} is used to minimise the gap between the source and target domains. In contrast to these studies, ADDi utilises a dual discriminator based network to learn a domain invariant representation by bringing source and target features closer to each other with three-players adversarial minimax games hence producing better results. Using the ablation study in subsection \ref{sub_ablation}, we further quantify the relative significance of our dual discriminator based approach.

Table \ref{crosscorpus} also shows the self-supervised learning (SSL) for ADDi which we called sADDi above. When we pre-train the encoder component in the sADDi network using the proposed synthetic data generation pretext task, it learns to encode discriminative representation for synthetic emotional data generation through the process of accomplishing the proposed pretext task. This helps to produce emotionally discriminative domain generalised features while fine-tuning the encoder in sADDi and the baseline CNN. Results by the SSL methods are separated with a bold line in Table \ref{crosscorpus}, which shows that the pre-training of the encoder considerably improves the cross-corpus SER. It is worth noting that the performance of the baseline CNN is also improved by utilising the pre-trained encoder by our proposed pretext task, which attests the effectiveness of the proposed self-supervised pretext task.

\subsection{Cross-Language Results}
We evaluate ADDi and sADDi for cross-language SER using both dimensional and categorical emotions. We use the IEMOCAP and RECOLA datasets for dimensional emotion and perform binary arousal and valence classification. All data from one language is used as a training set and all samples of the respective target language are used as the test set. This is the same evaluation strategy used in \cite{neumann2018cross}. We also implement domain adaptive models, including the GAN and DANN for comparison on IEMOCAP and RECOLA. Cross-language SER results using IEMOCAP and RECOLA are presented in Table \ref{crosslanguage}. 
\begin{table}[!ht]
\centering
\scriptsize
\caption{Dimensional cross-language SER results by UAR (\%) using the IEMOCAP and RECOLA datasets.}
\begin{tabular}{|l|l|l|l|l|}
\hline
\multirow{2}{*}{Model} & \multicolumn{2}{c|}{\begin{tabular}[c]{@{}c@{}}IEMOCAP (\tiny{English}) \\ to RECOLA (\tiny{French})\end{tabular}} & \multicolumn{2}{c|}{\begin{tabular}[c]{@{}c@{}}RECOLA (\tiny{French})\\ to IEMOCAP (\tiny{English})\end{tabular}} \\ \cline{2-5} 
                       & arousal                                 & valence                                 & arousal                                 & valence                                \\ \hline
CNN \scriptsize{(baseline)} & 59.2$\pm$ 1.8         & 48.5$\pm$ 1.5      & 60.7$\pm$ 1.6       & 48.3$\pm$  2.0          \\ \hline
ACNN \cite{neumann2018cross}&59.3&49.1&61.2&47.5\\\hline
GAN \cite{latif2019unsupervised}&59.8$\pm$1.9&49.8$\pm$1.7&60.3$\pm$1.3&48.7$\pm$ 1.5\\\hline
DANN \cite{abdelwahab2018domain}&60.1$\pm$2.1&50.2$\pm$1.5&61.5$\pm$1.5&49.2$\pm$ 1.4\\\hline
\rowcolor[HTML]{EFEFEF}
ADDi \scriptsize{(proposed)}   & \textbf{61.5$\pm$1.2}    & \textbf{51.8$\pm$1.4}        & \textbf{62.2$\pm$1.3}       &\textbf{50.9$\pm$1.2}                                  \\ \hlineB{3}\hlineB{1}
CNN$_\text{SSL}$  \scriptsize{(baseline)}      & 60.1$\pm$1.5       & 49.2$\pm$   1.3            & 61.2 $\pm$   1.6        & 49.0$\pm$   1.4      \\ \hline
\rowcolor[HTML]{EFEFEF}
sADDi \scriptsize{(proposed)}       & \textbf{63.8$\pm$1.0}         &\textbf{ 53.8$\pm$1.2}                   &\textbf{64.2$\pm$1.4}                   & \textbf{52.5$\pm$1.3}                              \\   \hline
\multicolumn{5}{|c|}{using fraction of target date for fine-tuning.} \\ \hline
ACNN \cite{neumann2018cross} (\tiny{500 target samples})&67.03&50.42&63.07&49.81\\\hline

\rowcolor[HTML]{EFEFEF}
sADDi \tiny{(250 target samples)}       & \textbf{70.3$\pm$1.3}         &\textbf{57.1$\pm$1.3}                   &\textbf{68.6$\pm$1.0}                   & \textbf{56.3$\pm$1.1}                              \\    \hline
\multicolumn{5}{|c|}{using language information.} \\ \hline

\rowcolor[HTML]{EFEFEF}
sADDi \tiny{(250 target samples) }     & \textbf{72.4$\pm$1.6}         &\textbf{60.1$\pm$1.4}                   &\textbf{70.2$\pm$1.3}                   & \textbf{59.3$\pm$1.2}      \\ \hlineB{3}\hlineB{1} 
\end{tabular}
\label{crosslanguage}
\end{table}

Using our proposed ADDi framework, we achieve better results compared to \cite{neumann2018cross}, where the authors use an Attentive Convolutional Neural Network (ACNN) to achieve promising results by fine-tuning the model on the target language. We also compare our results with domain adaptation architectures, including GAN and DANN in Table \ref{crosslanguage}. Compared to these models, ADDi is able to capture an emotion discriminative generalised representation by adversarially minimising the domain shift among source and target language data to improve SER across different language data. Performance is further improved when features learnt through SSL are utilised to guide the cross-language domain adaptation using sADDi. It is important to note that the performance of all the models is close to the chance level UAR (i.e., 50 \%), which shows the complexity of cross-language SER. However, our model improves the baseline results compared to the previous studies. To further improve the baseline performance, we perform two experiments. In the first experiment, we utilise a fraction of target data in the training set. Results in Table \ref{crosslanguage} show that including only 250 target language data yields considerable improvements compared to ACNN \cite{neumann2018cross} with 500 target samples. We incorporate language id with the source language data and 250 target language samples in the second experiment. Results are reported in Table \ref{crosslanguage}, which shows that the performance is improved for both arousal and valance prediction using the language information in the training data. However, the language information helps valence prediction more than the arousal prediction, which indicates that valence is more lexically dependent than arousal. 

We also compare our results on categorical cross-language emotion classification with different studies \cite{latif2018transfer,parry2019analysis,deng2013sparse,latif2019unsupervised,abdelwahab2018domain} and present the results in Table~\ref{crosscategorical}. In \cite{latif2018transfer}, the authors use transfer learning to improve cross-language SER using DBNs. A CNN-LSTM is suggested in \cite{parry2019analysis} and an autoencoder is tested in \cite{deng2013sparse} for cross-language SER. We also compare our results with GAN and DANN-based domain adaptive implementations for cross-language SER. ADDi achieves better results compared to all, which attests that ADDi learns greater domain generalised representation for cross-language scenarios. Compared to baseline, ADDi achieves 3.9 \% and 2.8 \% relative improvements for IEMOCAP to EMODB and EMODB to IEMOCAP experiments, respectively. Similar to the dimensional emotions, performance on categorical cross-language SER is further boosted by using a self-supervised ADDi (sADDi) network, which shows that the network is able to produce better generalised features for cross-language SER guided by the synthetic emotional data generation pre-training.

\begin{table}[!ht]
\caption{Categorical cross-language SER results by UAR (\%) using the IEMOCAP and EMODB datasets.}
\centering
\scriptsize
\begin{tabular}{|l|l|l|}
\hline
Model    & \begin{tabular}[c]{@{}l@{}}IEMOCAP (\tiny{English})\\  to EMODB (\tiny{German})\end{tabular} & \begin{tabular}[c]{@{}l@{}}EMODB (\tiny{German}) \\ to IEMOCAP (\tiny{English})\end{tabular} \\ \hline
CNN \scriptsize{(baseline)}    & 42.2$\pm$ 1.9      &  38.4$\pm$2.2    \\ \hline
DBN \cite{latif2018transfer}  & 42.5$\pm$2.1    &        39.5$\pm$2.4                                                \\ \hline
CNN-LSTM \cite{parry2019analysis}&42.1$\pm$1.8&38.9$\pm$2.1\\\hline
AE \cite{deng2013sparse}& 43.2$\pm$2.3    &        40.1$\pm$1.8                                              \\ \hline
GAN \cite{latif2019unsupervised}&44.3$\pm$1.7&40.3$\pm$1.7\\\hline
DANN \cite{abdelwahab2018domain}&43.5$\pm$1.8&40.5$\pm$2.0\\\hline
\rowcolor[HTML]{EFEFEF}
ADDi \scriptsize{(proposed)} &  \textbf{46.1$\pm$1.6}      &   \textbf{41.2$\pm$1.8}                                                          \\ \hlineB{3}\hlineB{1}
CNN$_\text{SSL}$  \scriptsize{(baseline)}     &   43.5$\pm$1.7      &   40.2$\pm$1.9   \\ \hline
\rowcolor[HTML]{EFEFEF}
sADDi \scriptsize{(proposed)} &  \textbf{48.3$\pm$1.5}       &   \textbf{44.8$\pm$1.6}    \\ \hlineB{3}\hlineB{1}
\end{tabular}
\label{crosscategorical}
\end{table}


\subsection{Impact of pretext selection: Reconstruction versus Synthetic data generation}
\label{sec:selfsup}
We propose synthetic data generation as a pretext task for self-supervised learning. In this experiment, we evaluate the effectiveness of this pretext task by comparing it with reconstruction as a pretext task. We make the comparison for both, the baseline and the ADDi networks.
 
 Reconstruction is widely used as a pretext task in the computer vision literature \cite{jing2020self,shukla2020visual}, wherein an autoencoder network is used to reconstruct the input back from the compressed representation. To use reconstruction as a pretext, we use unlabelled data (LibriSpeech) for unsupervised reconstruction and pre-train the encoder component to be utilised in the downstream task.
 We use the LibriSpeech data to generate the synthetic emotional data and pre-train the encoder component to use synthetic data generation for pretext. 
 
\begin{figure*}[!t]%
\centering
\begin{subfigure}{0.45\linewidth}
\includegraphics[width=\linewidth]{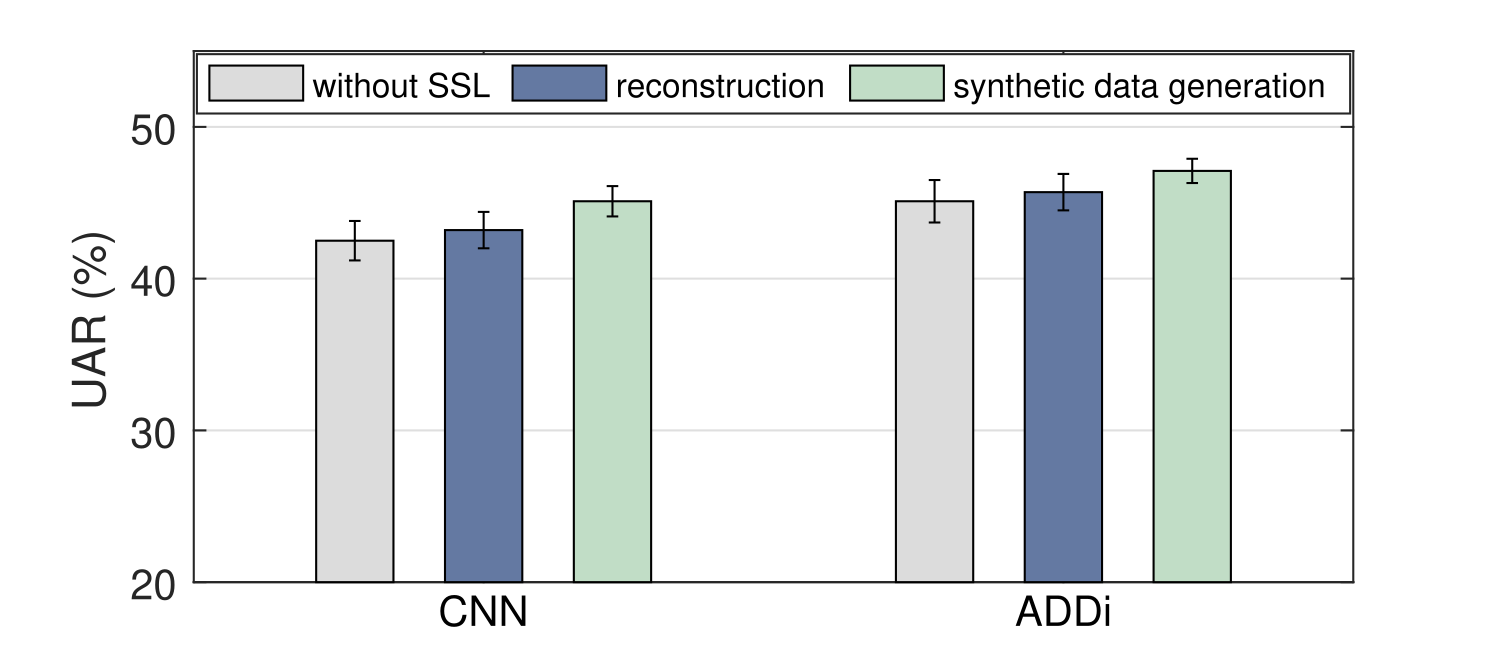}%
\captionsetup{justification=centering}
\caption{IEMOCAP to MSP-IMPROV}%
\label{IEMMSPSSL}%
\end{subfigure}\hfill%
\begin{subfigure}{0.45\linewidth}
\includegraphics[width=\linewidth]{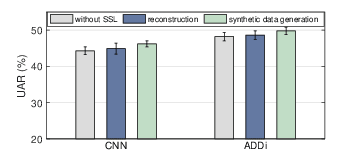}%
\captionsetup{justification=centering}
\caption{MSP-IMPROVt to IEMOCAP} %
\label{MSP2IEMOSSL}%
\end{subfigure}\hfill

\begin{subfigure}{0.45\linewidth}
\includegraphics[trim=0.1cm 0.1cm 0.1cm 0.1cm,clip=true,width=\linewidth]{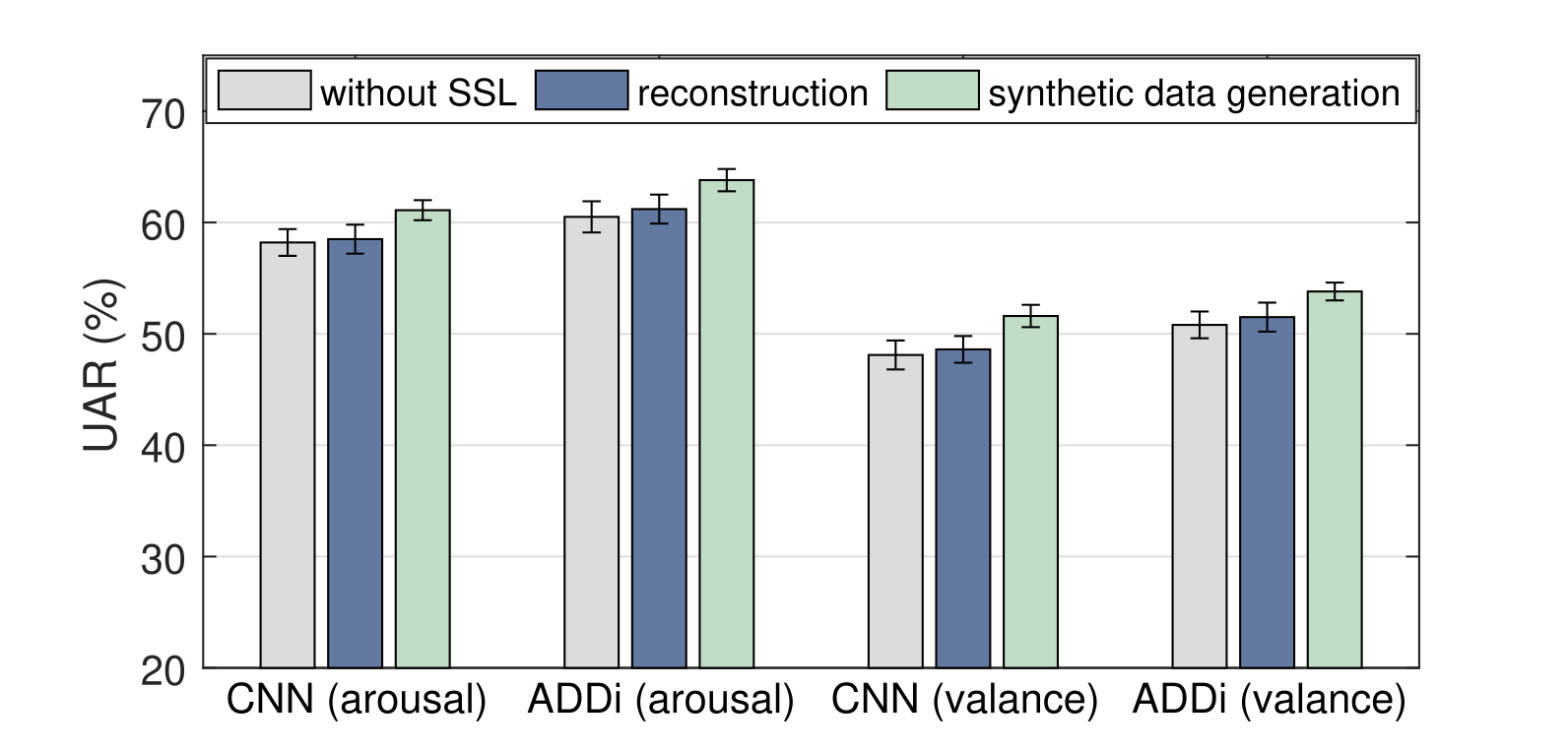}%
\captionsetup{justification=centering}
\caption{IEMOCAP to RECOLA}%
\label{IEM2RECSSL}%
\end{subfigure}\hfill%
\begin{subfigure}{0.45\linewidth}
\includegraphics[trim=0.0cm 0.1cm 0.1cm 0.1cm,clip=true,width=\linewidth]{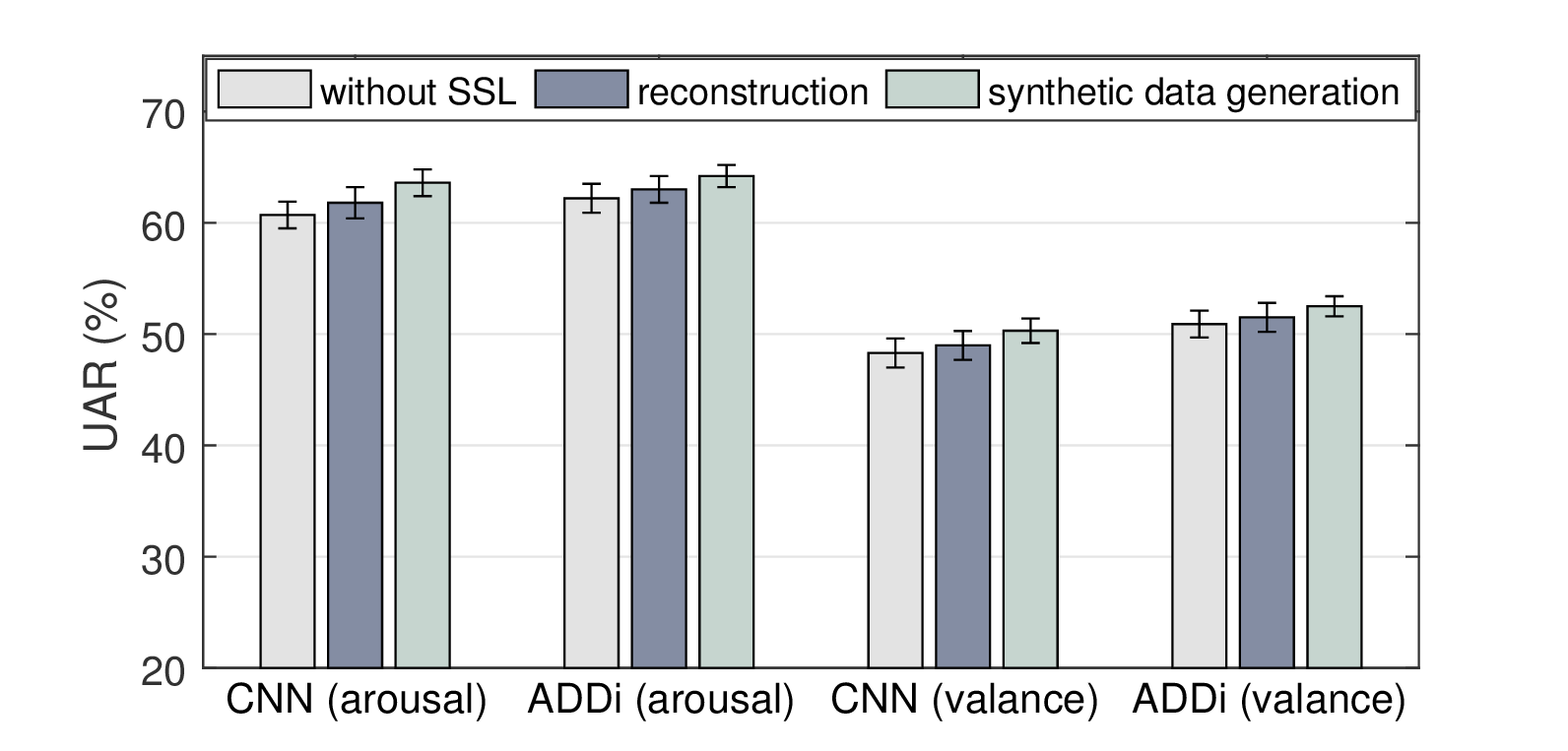}%
\captionsetup{justification=centering}
\caption{RECOLA to IEMOCAP} %
\label{REC2IEMOSSL}%
\end{subfigure}%
\caption{Impact of self-supervised pre-training on cross-corpus SER (Figure \ref{IEMMSPSSL} and \ref{MSP2IEMOSSL}) using the IEMOCAP and MSP-IMPROV datasets and cross-language SER (Figure \ref{IEM2RECSSL} and \ref{REC2IEMOSSL}) using IEMOCAP and RECOLA.}.
\label{fig:crosscorpusSSL}
\end{figure*}

 Results of the comparisons are presented in Figure \ref{fig:crosscorpusSSL} for cross-corpus and cross-language SER. For cross-corpus SER, we use IEMOCAP and MSO-IMPROV, and IEMOCAP and RECOLA are used for cross-language SER.
Compared to the reconstruction-based pretext task, we achieve better results using synthetic emotional data generation for both the baseline CNN model and ADDi. However, despite the popularity of autoencoder-based reconstruction pretext tasks in computer vision, it could not produce strong representations for transfer tasks in SER. One possible reason might be that the autoencoder only learns to encode abstract bottleneck representations from non-emotional speech, which cannot provide supervisory signals in the downstream domain adaptation task. 




\subsection{Impact of Data Augmentation}
\label{sec:syndata}

We generate the synthetic data during our pre-training step and use it to augment the source training data. We evaluate the model in a cross-corpus setting using IEMOCAP as training and MSP-IMPROV as testing data to compare the results with \cite{sahu2018enhancing,bao2019cyclegan,latif2020augmenting}. In \cite{sahu2018enhancing}, the authors investigate a GAN to generate the synthetic feature vectors using low dimensional features to augment the SER. Bao et al.\ \cite{bao2019cyclegan} apply a CycleGAN based model for synthetic samples by transferring feature vectors extracted from a large unlabelled speech data into the target synthetic emotional samples. They augment the SER system with synthetic features to improve SER performance. Recently, Latif et al.\ \cite{latif2020augmenting} utilise the combination of a GAN and mixup \cite{zhang2018mixup} to generate synthetic samples for SER augmentation. Similar to these studies, we also augment the SER system with synthetic data and perform evaluations using real, synthetic, and real plus synthetic data. We also use MSP-IMPROV as the target data, as per these studies. We randomly select 30$\,\%$ of the data as a development set for hyper-parameter selection and the remaining 70$\,\%$ as testing data as used in these studies. We keep these splits speaker independent. 
Results are compared with these studies in Table \ref{syntheticdata}. As expected, the synthetic data alone cannot offer better results, but we get better performance when we augment source data with the synthetic data to train ADDi.

\begin{table}[!ht]
\caption{UAR comparison for cross-corpus evaluation using synthetic data to augment the model. }
\begin{tabular}{|l|l|l|l|}
\hline 
Studies              & Real         & Syn.         & Real+Syn    \\ \hline
Sahu et al. \cite{sahu2018enhancing} & 45.14        & 33.96        & 45.40       \\ \hline
Bao et al. \cite{bao2019cyclegan}  & 45.58 $\pm$ 0.40 & 41.58 $\pm$  1.29 & 46.5$\pm$ 0.43  \\ \hline
Latif et al.  \cite{latif2020augmenting}       & 46.0$\pm$ 0.57    & 42.15 $\pm$ 1.12 & 46.60 $\pm$ 0.45 \\ \hline
\rowcolor[HTML]{EFEFEF}
ADDi             &  \textbf{47.83$\pm$ 0.45 }           & \textbf{ 42.25 $\pm$ 0.95}           & \textbf{48.61    $\pm$ 0.40   }     \\ \hline
\end{tabular}
\label{syntheticdata}
\end{table}

\subsection{Impact of incorporation of Source/Target Data}
\begin{figure*}[!ht]%
\centering
\begin{subfigure}{0.49\linewidth}
\includegraphics[trim=0cm 0cm 0cm 0cm,clip=true,width=\linewidth]{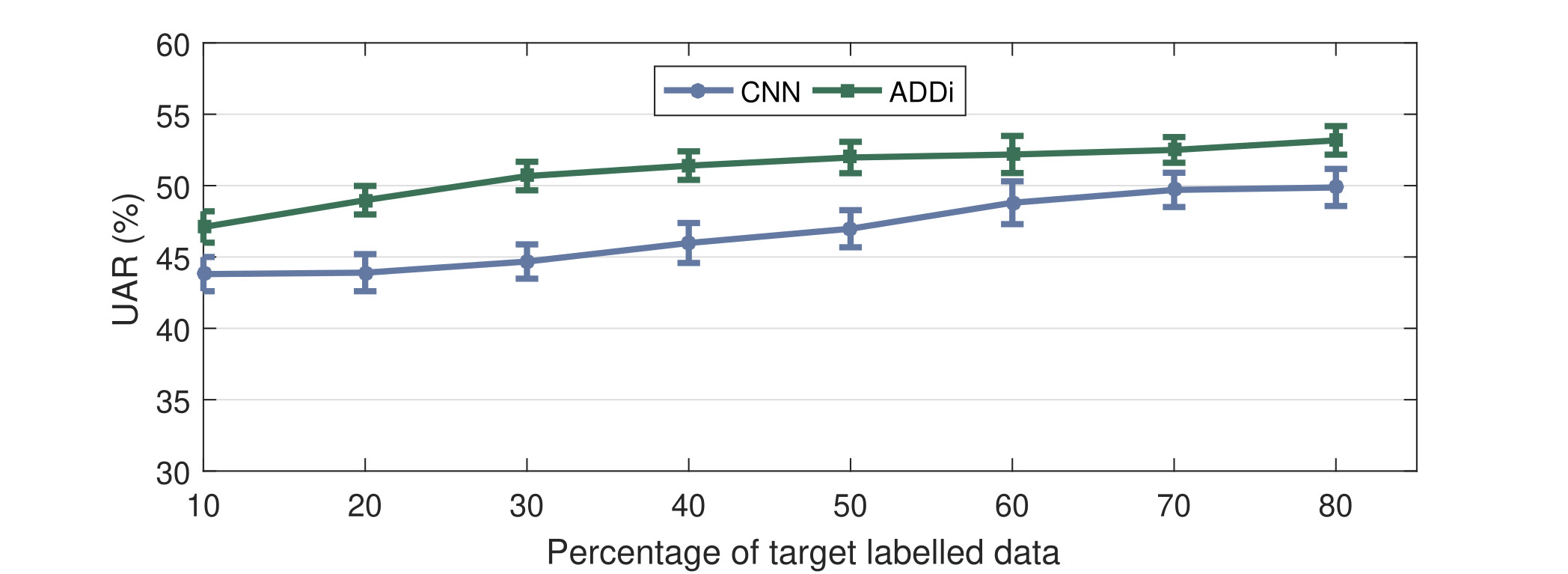}%
\captionsetup{justification=centering}
\caption{IEMOCAP to MSP-IMPROV}%
\label{IEMMSPPER}%
\end{subfigure}\hfill%
\begin{subfigure}{0.49\linewidth}
\includegraphics[trim=0.0cm 0cm 0cm 0cm,clip=true,width=\linewidth]{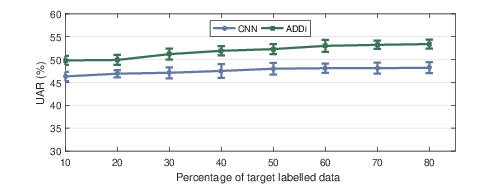}%
\captionsetup{justification=centering}
\caption{MSP-IMPROV to IEMOCAP} %
\label{MSP2IEMOPER}%
\end{subfigure}\hfill
\begin{subfigure}{0.49\linewidth}
\includegraphics[trim=0cm 0cm 0cm 0cm,clip=true,width=\linewidth]{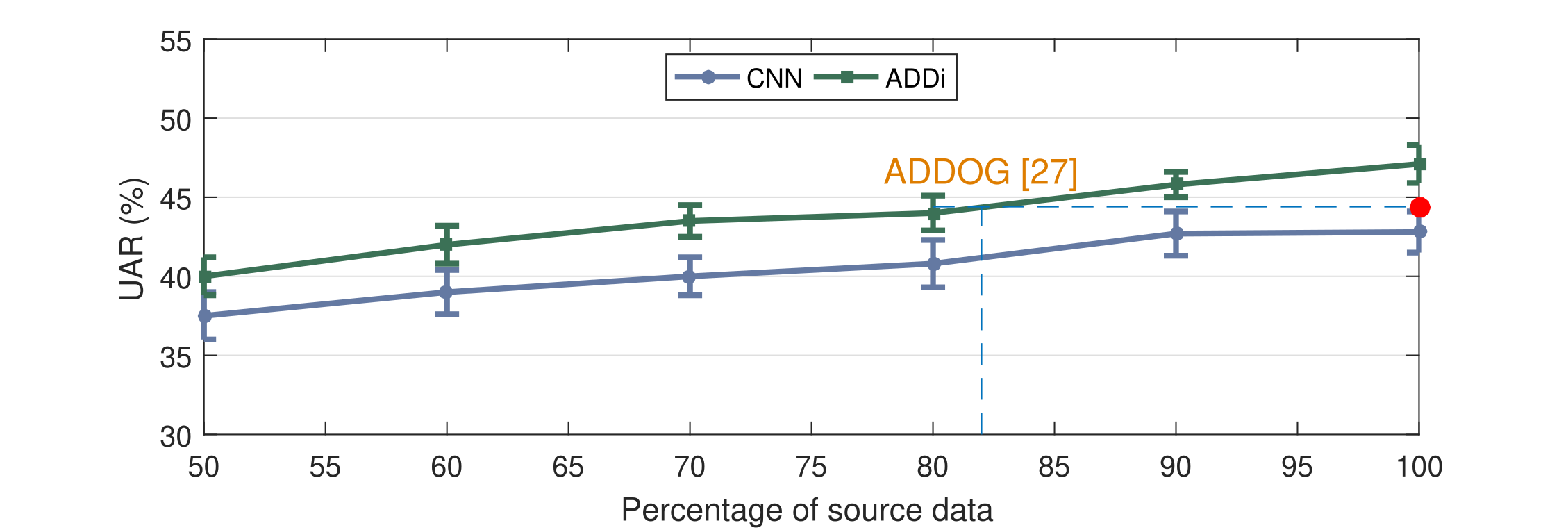}%
\captionsetup{justification=centering}
\caption{IEMOCAP to MSP-IMPROV}%
\label{IEMMSPPER_S}%
\end{subfigure}\hfill%
\begin{subfigure}{0.49\linewidth}
\includegraphics[trim=0.0cm 0cm 0cm 0cm,clip=true,width=\linewidth]{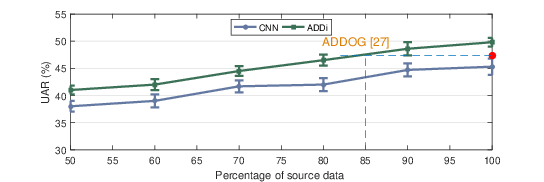}%
\captionsetup{justification=centering}
\caption{MSP-IMPROV to IEMOCAP} %
\label{MSP2IEMOPER_S}%
\end{subfigure}%
\caption{Results for cross-corpus SER with increasing amounts of labels from the target and source datasets.}
\label{fig:per}
\end{figure*}

This experiment incorporates the labelled target data into the training and validation. Here, we present the results using IEMOCAP and MSP-IMPROV in Figure \ref{fig:per}. Similar results are achieved for cross-language datasets. We plot the results with different percentages of target data using the baseline 
approach 
and ADDi. ADDi improves the results considerably against baseline CNN in all the case, even for a small percentage target labelled data. Figure \ref{fig:per} shows that the margin of UAR improvement decreases with incorporating larger percentages of labelled target data. This may indicate that the generalisation effect diminishes once there is sufficient amount of labelled target domain data.

We also explore the effect of decreasing the percentage of source data on the performance of cross-corpus SER using IEMOCAP and MSP-IMPROV. In both cases, sADDi performs better than the baseline. We also compare the results with ADDOG \cite{gideon2019improving} in Figure \ref{IEMMSPPER_S} and \ref{MSP2IEMOPER_S}. The red dot shows the performance achieved by ADDOG using 100 per cent of source data. We achieve these results using 80-86\,\% of source data as highlighted by a dotted blue line.

\begin{table}[!ht]
\centering
\caption{Results with the naturalist speech using FAU-AIBO corpus.}
\begin{tabular}{|l|l|l|}
\hline
Model    & \begin{tabular}[c]{@{}l@{}}IEMOCAP (\tiny{English})\\ to FAU-AIBO (\tiny{German})\end{tabular} & \begin{tabular}[c]{@{}l@{}}FAU-AIBO (\tiny{German})\\ to IEMOCAP (\tiny{English})\end{tabular} \\ \hline
CNN (\tiny{baseline}) &     53.7$\pm$1.8                          &       52.1$\pm$1.5                                                        \\ \hline
DBN  \cite{latif2018transfer}    & 54.5  $\pm$2.0                                                  &    50.7 $\pm$1.7                                                        \\ \hline
sADDi    &     \textbf{ 58.3  $\pm$ 1.5       }                                               &  \textbf{56.9  $\pm$1.6        }                                                   \\ \hline
\end{tabular}
\label{fau}
\end{table}
\subsection{Evaluations in The Wild}
In this section, we evaluate the performance of the proposed model on the naturalist speech. For this experiment, we use FAU-AIBO corpus that contains the natural speech of children in the German language. We perform binary valence classification as used by previous studies \cite{latif2018transfer,eyben2015geneva}. This experiment is more difficult compared to the previous experiments, due to the difference in language, age, and elicitation strategy. We train the model on source data and 20\% target data is used as validation and the remaining is used for testing. Results for both experiments are reported in Table \ref{fau}, which shows that the proposed model considerably improved performance compared to the DBNs \cite{latif2018transfer} and baseline CNNs. sADDi is improving the results by above 4\% for both experiments presented in Table \ref{fau}.  This confirms the effectiveness of our sAADi network that can produce generalised representations for evaluations against naturalist speech.

\subsection{Size of Pretext Task Training Data}

\begin{figure}[!t]%
\centering
\includegraphics[trim=0cm 0cm 0cm 0cm,clip=true,width=1\linewidth]{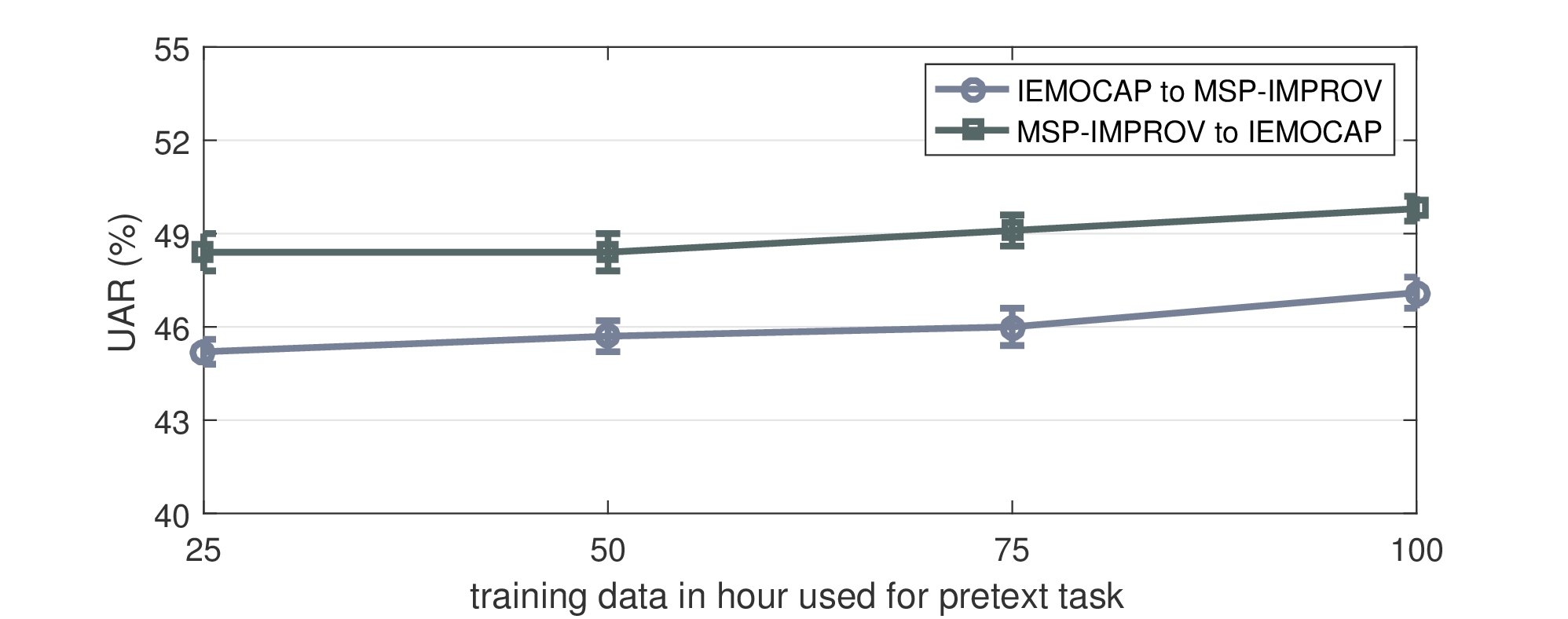}%
\captionsetup{justification=centering}
\caption{Effect of increasing the training data in hours on the performance (UAR \%). }%
\label{pretext}%
\end{figure}

We next examine the effect of the size of training data in pretext tasks on the performance of sADDi. We plot the results for IEMOCAP and MSP-IMPROV in Figure \ref{pretext}. For both experiments, we find that the increase of training data in the pretext task helps improve the performance of the downstream emotion classification. This shows that increasing the training data in the pretext task enables the model to produce representations suitable for the downstream task of speech emotion recognition (SER).

\subsection{Ablation Experiments}
\label{sub_ablation}

In this experiment, we validate the necessity and effectiveness of each module integrated with our proposed model ADDi. Results are presented in Table \ref{ablation} for cross-corpus and cross-language evaluations using IEMOCAP to MSP-IMPROV and IEMOCAP to EMODB. These results are computed without any data augmentation and pre-training. Results for synthetic data augmentation and pre-training are presented in Section \ref{sec:syndata} and \ref{sec:selfsup}. This experiment starts with the ADDi model (model 1) that contains all the components, including the encoder, two discriminators, the generator/decoder, and the classifier. We remove one discriminator in models 2 and 3. This makes the model similar to the standard GAN with an additional classifier and autoencoder. We keep removing different modules until we obtain the baseline CNN network (model 5), containing only the encoder and classifier. We also plot the configurations of these models 1-5 in Table \ref{ablation}. There is a 
considerable 
drop in UAR when one or more components are removed. When a single discriminator -- either $D_s$ or $D_t$ -- is used in model 2 or 3, we see a performance drop for both cross-corpus and cross-language SER. This shows that the single discriminator networks cannot achieve better generalisation compared to the three-players adversarial learning performed by the dual discriminator and generator approach in the ADDi network. Similarly,  ADDi also achieves 
considerably 
improved results compared to models 4 and 5 in cross-corpus and cross-language settings. In models 4 and 5, there is no component (i.\,e., discriminator) to promote generalised representations in the network by minimising the domain gap between source and target data. This shows that implanting a domain adaptation component in the pipeline of the deep model is important to learn to improve generalised features for cross-corpus and cross-language SER. Overall, these ablation experiments show that all the components in the proposed ADDi models are chosen carefully for effective domain adaptation for SER.

\begin{table*}[]
\centering
\scriptsize
\caption{Results for cross-corpus and cross-language using IEMOCAP to MSP-IMPROV, and IEMOCAP to EMODB, respectively. }
\begin{tabular}{|l|c|c|c|c|c|c|c|c|}
\hline
Model &Configuration & \multicolumn{1}{c|}{\begin{tabular}[c]{@{}c@{}}Discriminators\end{tabular}} & \multicolumn{1}{c|}{\begin{tabular}[c]{@{}c@{}}Decoder\end{tabular}} & \begin{tabular}[c]{@{}l@{}}Encoder\end{tabular} & \begin{tabular}[c]{@{}l@{}}Classifier\end{tabular}  & \begin{tabular}[c]{@{}c@{}}Cross-corpus\\UAR (\%)\end{tabular}&\begin{tabular}[c]{@{}c@{}}Cross-Language \\UAR (\%)\end{tabular} \\ \hline

1& \begin{minipage}{.3\textwidth}
      \includegraphics[width=\linewidth,height=23mm]{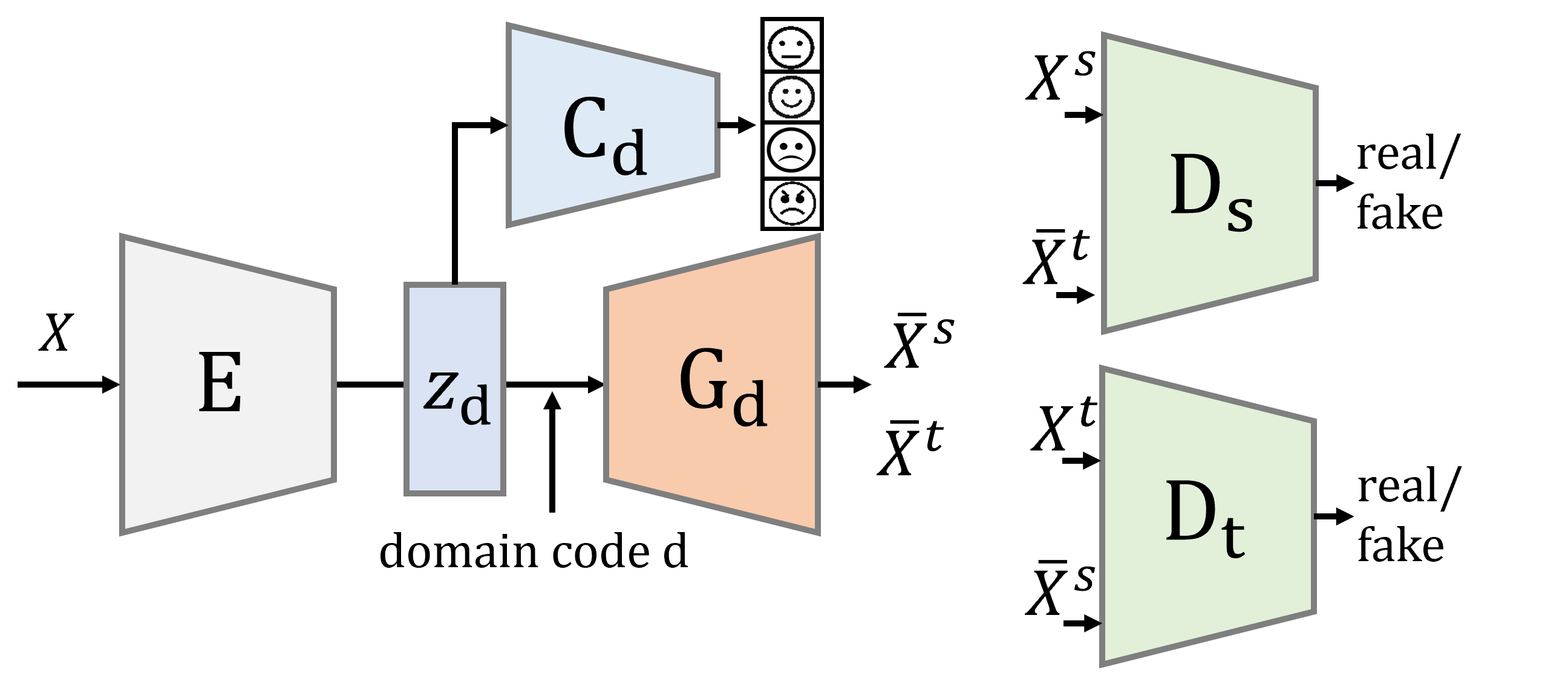}
    \end{minipage} & 2         &   \checkmark{}  &  \checkmark{}{}     &       \checkmark{}         &\textbf{45.1$\pm$0.8}      &\textbf{46.1$\pm$1.6}       \\ \hline
    2&\begin{minipage}{.3\textwidth}
      \includegraphics[width=\linewidth,height=23mm]{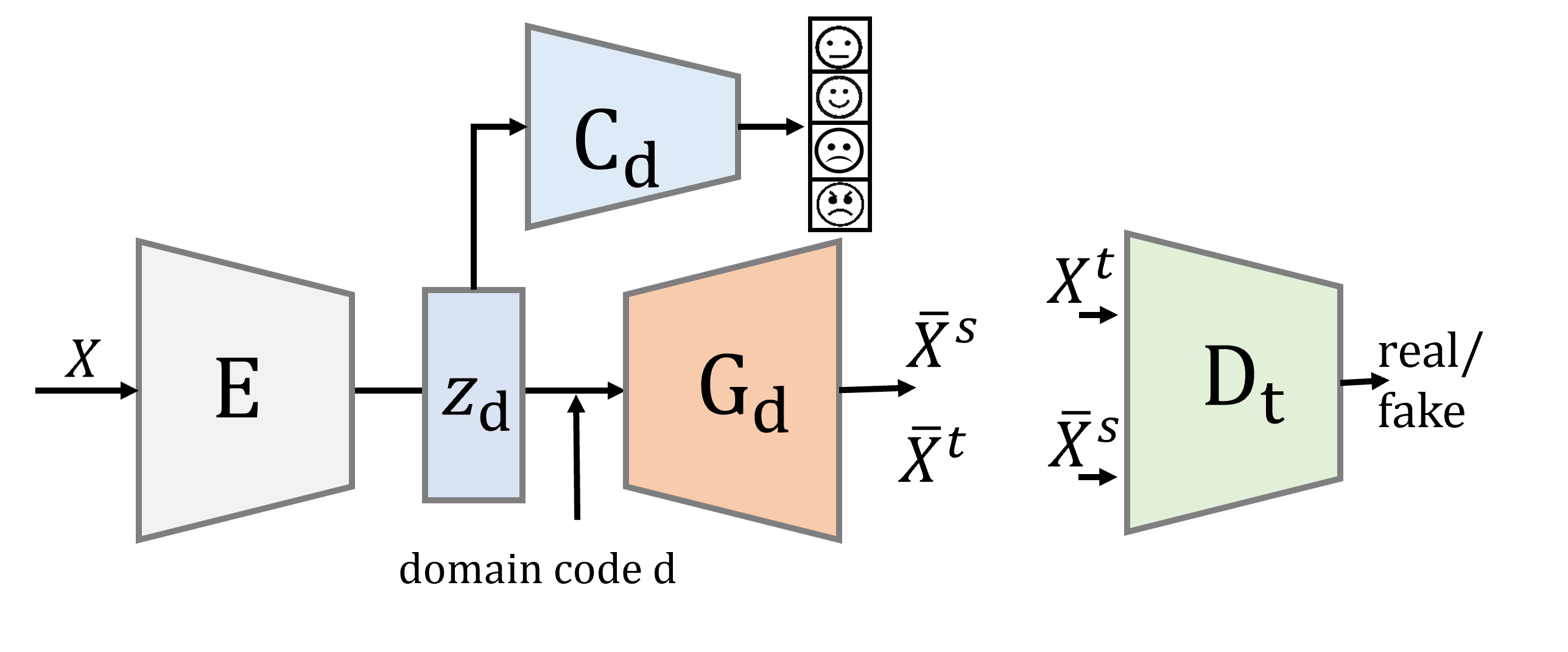}
    \end{minipage} &   1   &  \checkmark{}   &   \checkmark{}  &    \checkmark{}              &43.2$\pm$1.3       &44.0$\pm$1.9      \\ \hline

   3& \begin{minipage}{.3\textwidth}
      \includegraphics[width=\linewidth,height=23mm]{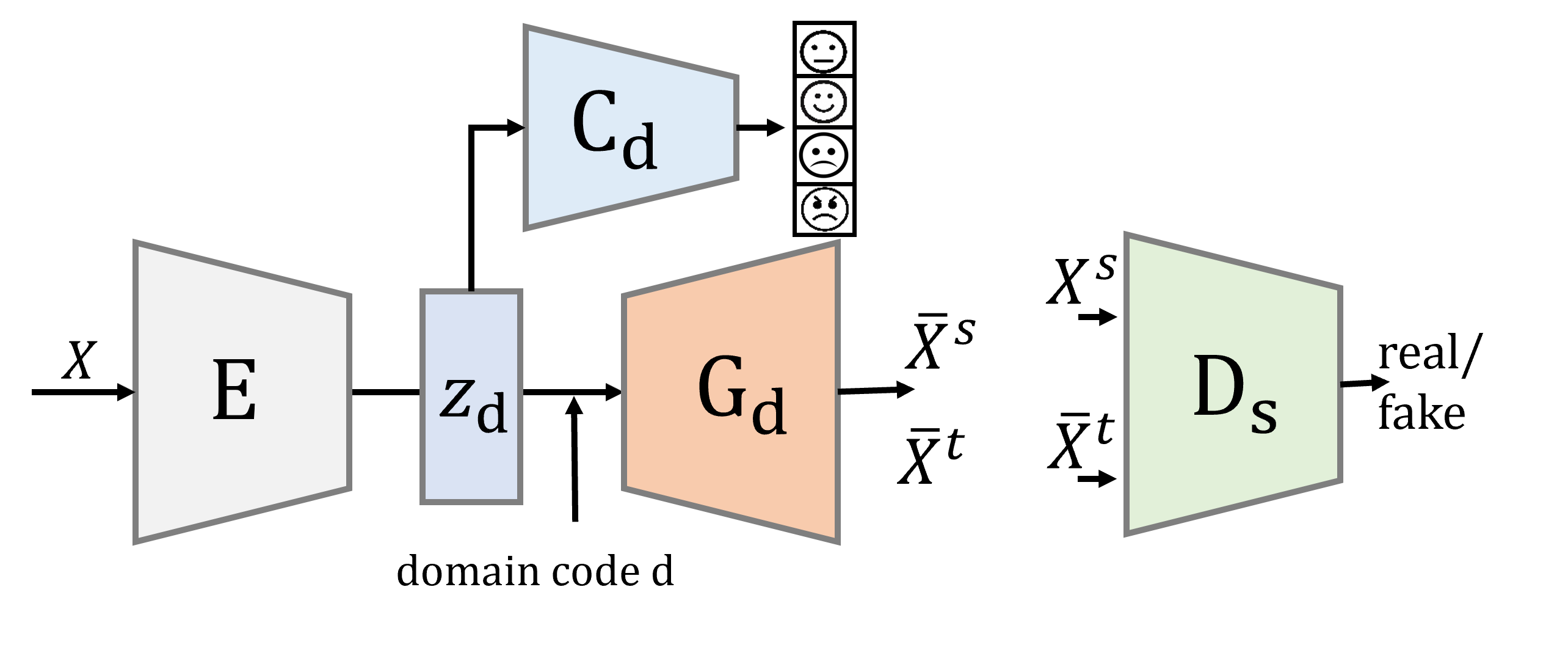}
    \end{minipage} &   1   &  \checkmark{}   &   \checkmark{}  &    \checkmark{}              &43.1$\pm$1.4      &43.5$\pm$1.8    \\ \hline
    4&\begin{minipage}{.3\textwidth}
      \includegraphics[width=\linewidth,height=23mm]{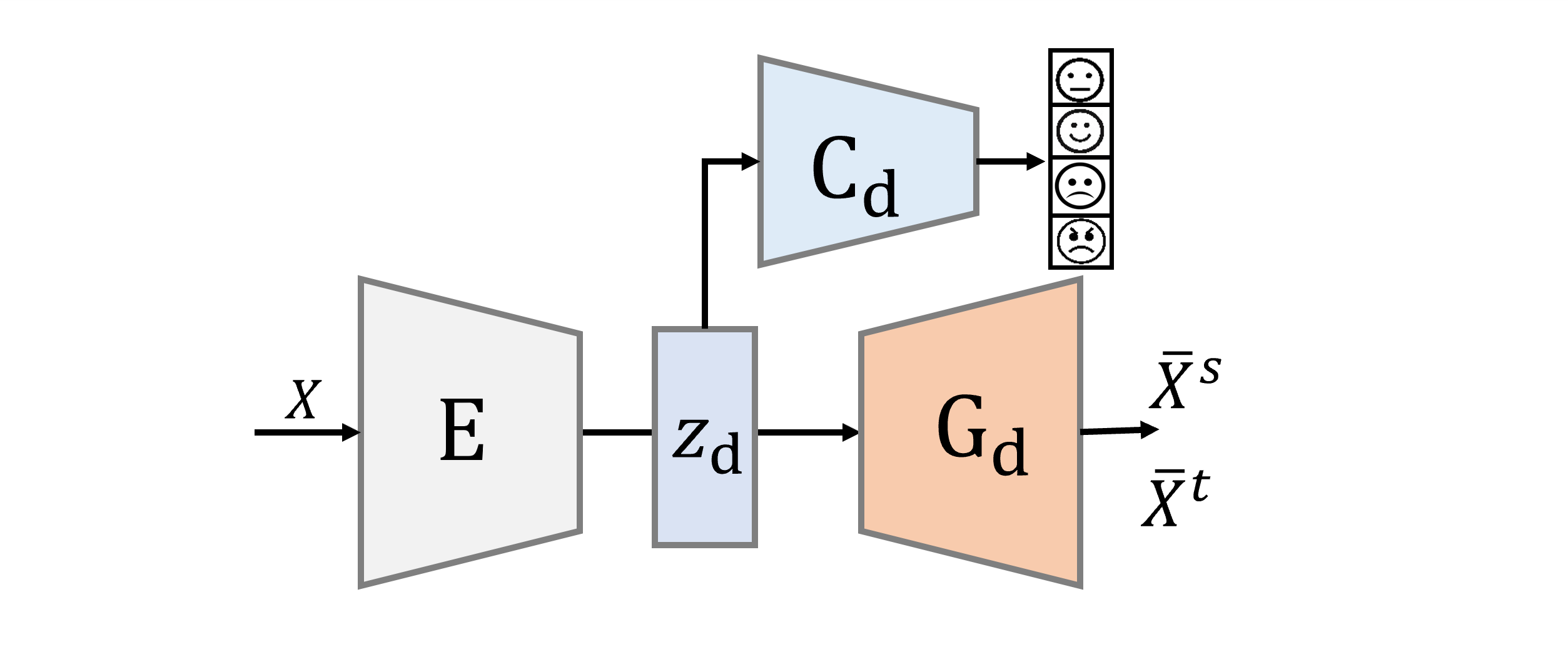}
    \end{minipage}  &  \xmark{}   &\checkmark{}       &     \checkmark{}                          & \checkmark{}          &   42.7$\pm$1.2     &43.3$\pm$1.8    \\ \hline

  5&\begin{minipage}{.3\textwidth}
      \includegraphics[width=\linewidth,height=23mm]{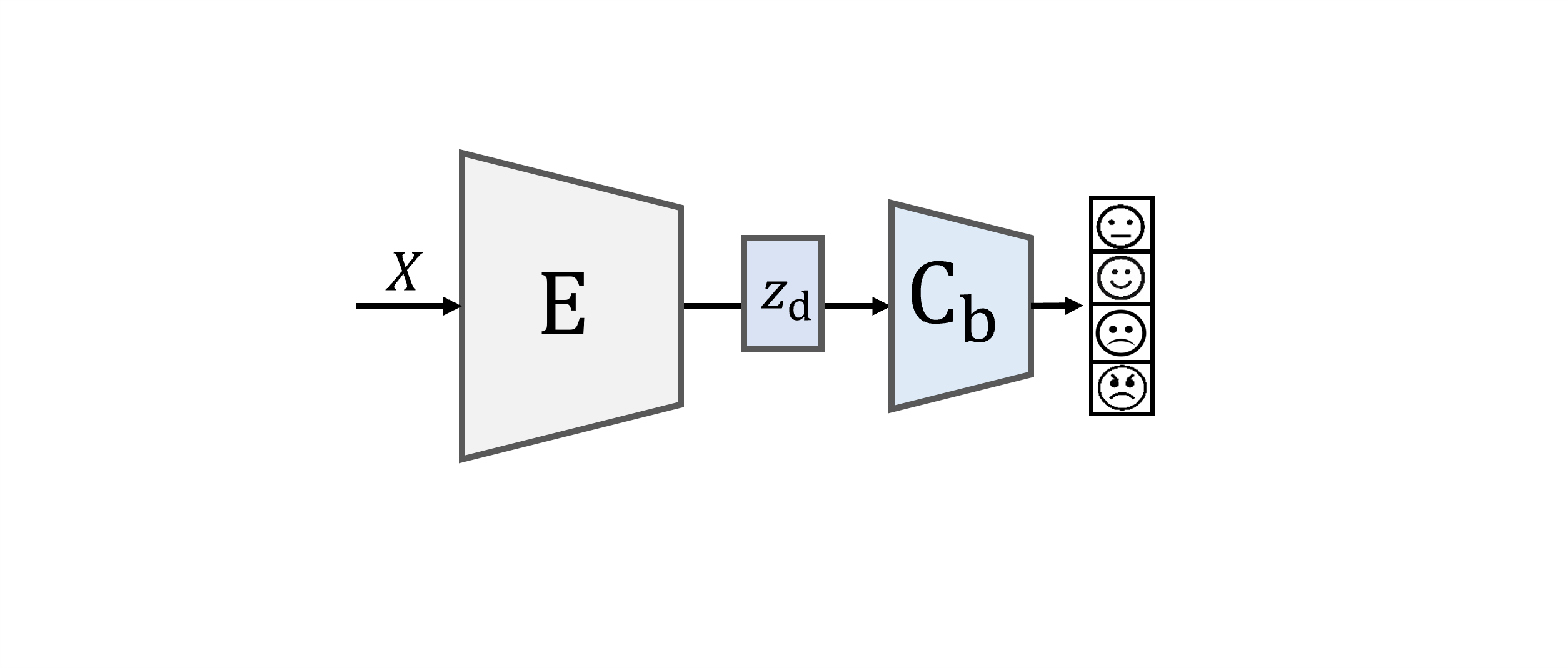}
    \end{minipage}  &     \xmark{}                   &  \xmark{}               &  \checkmark{}  
     &   \checkmark{}               & 42.5$\pm$1.6    &41.2$\pm$1.9     \\ \hline

\end{tabular}
\label{ablation}
\end{table*}

\section{Conclusions and Future Work}
This contribution addressed the open challenge of 
improving the speech emotion recognition (SER) performance in cross-corpus and cross-language settings. We proposed the Adversarial Dual Discriminator (ADDi) network that minimises the domain shift among emotional corpora adversarially. We focused on exploiting the unlabelled data with self-supervised pre-training and proposed self-supervised ADDi (sADDi). For sADDi, we suggested synthetic data generation as a pretext task, which (1) helped improve the domain generalisation performance of an SER system to tackle the larger domain shift between training and test distributions in cross-corpus and cross-language SER; and (2) produced byproduct synthetic emotional data to augment the SER system and minimise the requirement of source labelled data. The key highlights are as follows:

\begin{itemize}
    \item The introduced dual discriminator based ADDi network offers improved cross-corpus and cross-language SER without using any target data labels compared to the single discriminator and other state-of-the-art approaches. This is mainly due to the dual discriminator using a three-players adversarial game to learn generalised representations.  
    \item Considerable improvements in results were found when partial target labels were fed to the network training. This helped the ADDi to regulate the generalised representations based on the target data by maximally matching the data distributions.
    
  \item Our proposed self-supervised pretext task produces synthetic data as a byproduct to augment the system to achieve better performance. We were able to reduce 15-20$\,\%$ source training data using sADDi while achieving similar performance reported by a recent related study \cite{gideon2019improving}.
    \end{itemize}

Future studies will include evaluating the ADDi and sADDi architectures to model other factors of speech variations, including age, subject, gender, phoneme, noise, and recording device. Further experiments may include evaluating the proposed methods in wild conditions like noisy speech and adversarial noise. We are also interested in exploring multimodal pretext task techniques in our future work. Multimodal human interaction in video and textual form can provide various opportunities for self-supervised learning to improve cross-corpus and cross-language SER.

\end{document}